\documentclass{emulateapj}
\usepackage{hyperref}
\usepackage{graphicx}
\usepackage{color}
\usepackage{multirow}
\usepackage{amsmath}
\usepackage{rotating}
\usepackage{rotfloat}
\def\hideprivate{\global\long\def\private##1{\iffalse ##1 \fi}}
\hideprivate

\usepackage{ulem}
\def\Shf{S_h(f)}

\def\Pol{F^2 ({\bf \hat r},t)}
\def\Mchirp{{\cal M}}
\def\Pfac{\langle F^2\rangle_{{\bf \hat r},t}}

\def\Rbirth{\mathcal{R}_{\rm birth}(f,{\bf r})}
\def\Rmergef{\mathcal{R}_{\rm mrg}(f,{\bf r})}
\def\Rmrg{\mathcal{R}_{\rm mrg}({\bf r})}
\def\Rtot{\mathcal{R}_{\rm gal}}

\def\Deff{\mathcal{D}_{\rm char}}

\begin{document}

\title{Gravitational-wave emission from compact Galactic binaries}

\journalinfo{Astrophys. J. 758(131) (2012)}

\author{Samaya Nissanke,\altaffilmark{1,2} Michele Vallisneri,\altaffilmark{1,2} Gijs Nelemans,\altaffilmark{3,4,5} Thomas A. Prince\altaffilmark{1,2} }

\altaffiltext{1}{Jet Propulsion Laboratory,
	California Institute of Technology, 4800 Oak Grove Drive, Pasadena, CA 91109}
\altaffiltext{2}{Division of Physics, Mathematics, and Astronomy,
	California Institute of Technology, Pasadena, CA 91125}
\altaffiltext{3}{Department of Astrophysics, Radboud University Nijmegen,
	Heyendaalseweg 135, NL-6525 AJ Nijmegen, the Netherlands}
\altaffiltext{4}{Institute for Astronomy, KU Leuven, Celestijnenlaan 200D,
	3001 Leuven, Belgium} 
\altaffiltext{5}{Nikhef, Science Park 105, 1098 XG Amsterdam, The Netherlands} 

\shorttitle{GWs from compact Galactic binaries}
\shortauthors{Nissanke, Vallisneri, Nelemans, and Prince}

\begin{abstract}
Compact Galactic binaries where at least one member is a white dwarf
or neutron star constitute the majority of individually detectable
sources for future low-frequency space-based gravitational-wave (GW)
observatories; they also form an unresolved continuum, the dominant
Galactic foreground at frequencies below a few mHz. Due to the paucity
of electromagnetic observations, the majority of studies of
Galactic-binary populations so far have been based on
population-synthesis simulations. However, recent surveys have
reported several new detections of white-dwarf binaries, providing new
constraints for population estimates. In this article, we evaluate the
impact of revised local densities of interacting white-dwarf binaries
on future GW observations. Specifically: we consider five scenarios
that explain these densities with different assumptions on the
formation of interacting systems; we simulate corresponding
populations of detached and interacting white-dwarf binaries; we
estimate the number of individually detectable GW sources and the
magnitude of the confusion-noise foreground, as observed by
space-based detectors with 5- and 1-Mkm arms. We confirm earlier
estimates of thousands of detached-binary detections, but project only
few ten to few hundred detections of interacting systems. This
reduction is partly due to our assessment of detection prospects,
based on the iterative identification and subtraction of bright
sources with respect to both instrument and confusion noise. We also confirm earlier estimates for the confusion-noise foreground, except in one scenario that explains smaller local densities of interacting systems with smaller numbers of progenitor detached systems.
\end{abstract}

\keywords{gravitational waves, Galactic binaries, AM CVn, data analysis, LISA}

\section{Introduction}
\label{sec:intro}

Compact Galactic binaries with periods shorter than approximately five hours
are of considerable interest as gravitational wave (GW) sources (\citealt{Nelemans:2009,Marsh:2011}).
Such binaries include detached double white dwarf (DDWD) systems, short-period cataclysmic
variables (CVs), and AM CVn systems (so named after their prototype, AM Canum Venaticorum;
see \citealt{Solheim:2010}). These last are short-period binaries in
which a white dwarf (WD) is accreting He-rich material from an
H-deficient companion.  AM CVn systems are defined spectroscopically by the presence of strong
absorption or emission He lines and the absence of H lines, which
instead characterize ``standard'' CVs.  In this paper we focus on those binaries that are evolving toward, or have evolved from,
extremely short orbital periods, on the order of minutes.  These
include DDWD and AM CVn systems, sometimes referred to as ``ultra-compact'' binary systems.
Such binaries have received considerable attention, among other
reasons, as possible progenitors for Type Ia supernovae, R Coronae
Borealis stars \citep{Webbink:1984,Iben:1984}, and so-called ``.Ia'' explosions \citep{Bildsten:2007}. 
Our understanding of their formation and evolution remains poor. 

Together with ultra-compact X-ray binaries, DDWD and AM CVn systems will be the
most numerous and individually brightest GW sources in the Galaxy for
low-frequency  ($10^{-4}\mbox{--}0.1$ Hz) space-based GW
detectors. Within this frequency band, observations suggest $0.5 \mbox{--} 2.5 \times 10^7$
DDWD systems and theoretical estimates predict $10^5\mbox{--}10^7$ AM CVn systems in our Galaxy
\citep{Maxted:1999, NPVY01,Nelemans2001ddwd,Nelemans:2004}.  

The aim of this paper is to characterize the detectable GW emission from compact
binaries in light of recent population estimates based on new observations.
The first estimates were based solely on population-synthesis
simulations; see App.\ \ref{sec:prevwork} for a detailed discussion and review of
the literature. Recent observations of AM CVn systems have provided new information that constrains earlier estimates.
For instance, the identification of AM CVn systems among candidates
selected from the Sloan Digital Sky Survey (SDSS) spectroscopic
catalog allowed \cite{Roelofs:2007} to estimate the density of AM CVn                                                                                             
systems by modeling the completeness of the catalog.
A follow-up spectroscopic survey of a color-selected subset of the SDSS                                                                               
photometric database seems to find even fewer AM CVn systems                                                                              
than expected \citep{Roelofs:2007,Rau:2010}, leading to local density estimates 
smaller by $10\mbox{--}20$ than suggested by population-synthesis studies.

To date, 27 AM CVn systems have been presented in the literature; the most recent detections have been achieved in synoptic surveys, including a short-period system from Kepler \citep{Fontaine:2011}, and a medium-period, outbursting system from the Palomar Transient Factory \citep{Levitan:2011}.

Similarly, surveys of low-mass WD candidates selected from SDSS has
led to the discovery of several short-period DDWDs that could be
progenitors of AM CVn systems
\citep{Badenes:2009,Mullally:2009,Kulkarni:2010,Marsh:2010}.
For example, the ELM (Extremely Low Mass) survey
\citep{Brown:2010,Kilic:2011,Brown:2012} searches for
companions around extremely low mass WDs ($\sim 0.2 \, M_{\odot}$).
The ELM survey has approximately quintupled (with 19 new systems) the known sample of DDWDs that will merge within a Hubble time \citep{Kilic:2011b,Brown:2012},
providing evidence for at least one plausible AM CVn formation channel (see Sec.\ \ref{sec:ddwdchannel}).

In this paper, we consider five scenarios for AM CVn formation that span plausible populations, and we study their consequences for space-based GW detectors by modeling the detection and subtraction of sources in simulated data streams. 
We consider two representative detectors: the standard LISA developed as a joint NASA--ESA project until 2011 \citep{lisasciencecase,Jennrich:2009p1398}, and a descoped configuration with shorter armlengths of 1 Mkm, which is similar, but not equivalent, to the NGO/eLISA design studied by ESA in 2011 \citep{esastudy}.

We find, first, that the number of detectable AM CVn systems decreases from $\sim 10,000$, as reported in previous assessments of LISA science, to $\sim
100\mbox{--}400$ (for one year of observation with a single interferometer---see App.\ \ref{sec:interferometers}). Precise numbers depend on the assumptions used to explain how a
postulated number of DDWD progenitors can result in the
number of AM CVn systems seen in surveys.
For the shorter detector, these numbers are reduced by a factor of a few.
Second, we confirm that mostly DDWD systems determine the residual confusion foreground left after GW signals from detectable binaries are
subtracted. The foreground does not change significantly from previous
estimates, except in one scenario in which we explain the fewer observed AM CVn systems
by reducing the space density of DDWD progenitors.

The rest of this paper is structured as follows. In Sec.\
\ref{sec:formation} we discuss the current understanding of AM CVn
formation. In Sec.\ \ref{sec:revpop} we outline the observations of AM CVn and DDWD systems
in the Galaxy, and the resulting calibration of local space densities.
In Sec.\ \ref{sec:analysis} we present our analysis: we introduce our explanatory formation scenarios in Sec.\ \ref{sec:scenarios}, discuss the space-based observation of GWs from
Galactic binaries in Sec.\ \ref{sec:detecting},
describe our simulations of GW source detection and subtraction in Sec.\ \ref{sec:simulations}, and detail our results in Sec.\ \ref{sec:results}. In Sec.\ \ref{sec:conclusion} we recapitulate our conclusions and propose possible future work.  In the appendices we provide more detailed discussions of specific topics: in App.\ \ref{sec:prevwork} we summarize previous analytical and numerical studies of the Galactic foreground for LISA, and give a general semi-analytical derivation of its magnitude and shape; in App.\ \ref{sec:interferometers}
we discuss GW detection and define our reference detector configurations;
in App.\ \ref{app:subtracted} we provide more details about the fits
of the GW confusion foreground obtained from our simulations. In the
rest of the paper, we use the standard notation where $G$ is the
gravitational constant and $c$ is the speed of light.

\section{The formation and evolution of compact binary systems in the Galaxy}
\label{sec:formation}

In the standard picture of binary-star evolution, compact
binaries form following two consecutive mass-transfer episodes; the loss of sufficient angular momentum by friction in the
second common-envelope phase ensures sufficiently tight orbits
for GW emission to influence dynamics. Here we concentrate on the processes leading to mass transfer and to
the formation of AM CVn systems. Due to space limitations, we do not attempt here to provide an extensive literature review, and refer
the reader to \cite{Solheim:2010} and references therein for a detailed review on the current
understanding of the
formation and evolution of DDWD and AM CVn systems.

The subsequent orbital evolution of compact binaries is driven by gravitational
radiation, by mass transfer, by a dissipative coupling that can feed the angular
momentum added through accretion back into the orbit, and by other poorly
modeled processes such as wind loss, etc. Therefore, the rate of change
$\dot{J}_{\mathrm{orb}}$ of the orbital angular momentum comprises
three or more competing components,
\begin{equation}
\label{eqn:orbangmom}
\dot{J}_{\mathrm{orb}} = \dot{J}_{\mathrm{GR}} +
\dot{J}_{\mathrm{\dot{M}}} + \dot{J}_{\tau_s}.
\end{equation}
Here $\dot{J}_{\mathrm{GR}}$ represents the loss of angular momentum to gravitational radiation,
\begin{equation}
\label{eqn:angmomGW}
\dot{J}_{\mathrm{GR}} = -\frac{32}{5} \frac{G^3}{c^5} \frac{M_1 M_2 M}{a^4} J_{\mathrm{orb}}
\end{equation}
(for circular orbits) where $M_1$ and $M_2$ are the two binary masses
($M_2$ is the donor's mass for mass-transferring systems), $a$ their separation, and $M = M_1 + M_2$. Next, $\dot{J}_{\mathrm{\dot{M}}}=\sqrt{G M_1 R_h}
\dot{M}_2$, where $\dot{M}_2$ indicates the donor's mass-loss rate due to Roche-lobe overflow and $R_h$ defines the orbital radius around the accretor with the same specific angular momentum as the transferred mass (\citealt{Verbunt:1988}).

Mass transfer can proceed either via an accretion disk or via ``direct
impact'' (see, e.g., \citealt{MarshSteeghs:2002,Ramsay:2002,Marsh:2004}). ``Direct
impact'' occurs when the donor's mass stream hits the accreting star directly. The particular route depends on the
initial separation in the binary following the ejection of the second
envelope, and on the donor's entropy and equation of state
(see, e.g., \citealt{Gokhale:2007, Dan:2011, Dan:2012}). Typically,
mass-transferring binaries with initial orbital periods
greater than ten minutes form an accretion disk. As discussed below, the presence or absence of an accretion
disk is critical to the stability of the system. Whether a disk is
present or not, $\dot{J}_{\tau_s}$ denotes the change in the torque due to dissipative coupling, tidal or magnetic, between the accretor's spin
and the orbital angular momentum during the accretor's synchronization
timescale $\tau_s$. We assume that the donor and orbit's spins are synchronized. 
In this work, we do not consider the change in angular momentum driven by other physical processes,
such as losses due to winds or magnetic
braking, because of current limitations in our understanding and
modeling of such phenomena.

As the term ``detached'' in their very name implies, DDWD dynamics are
determined solely by gravitational radiation and tides (see e.g. \citealt{Piro:2011}). By contrast, the formation, stability, and outcome of AM CVn systems depend on all three effects in Eq.\ (\ref{eqn:orbangmom}). 
Three proposed channels exist for the formation of AM CVn systems, and each can vary in its efficiency. We describe them below. Following \cite{NPVY01,NYP01,Nelemans:2004}, we assume that the first two channels listed below can operate with either ``optimistic'' or ``pessimistic'' efficiencies, resulting in different AM CVn numbers and period distributions.
 
\subsection{The DDWD channel}
\label{sec:ddwdchannel}

A subclass of DDWDs with periods of several minutes becomes semi-detached when angular-momentum loss due to gravitational radiation
drives the two bodies sufficiently close that the lower-mass WD fills its Roche
lobe and mass transfer begins, either via direct impact or via an
accretion disk.  The stability of mass transfer depends critically on the mass ratio of
the two initial WDs (requiring typically $q = M_2/M_1 \lesssim 2/3$) and on whether spin--orbit tidal coupling feeds angular momentum accreted by the primary star back to the
orbit; the strength of this coupling is characterized by the 
synchronization timescale $\tau_s$ (\citealt{Marsh:2004}).

Under optimistic assumptions, strong tidal coupling stabilizes mass transfer, favoring the formation of stable AM CVn systems and resulting in fewer mergers. By contrast, under pessimistic assumptions there is no tidal coupling and many more DDWD systems
merge.  In the presence of a disk, the
accreted material's angular momentum is mainly transferred back into the orbit by tidal forces
at the outer edge of the disk. However, the disc's material at the
inner edge still has sufficient angular momentum to spin up the
accretor. Effects not considered here, such as magnetic braking (\citealt{Farmer:2010}), may
also impact the formation of AM CVn systems and favor the merger of the WDs.    

If stable mass transfer proceeds, the donor WD loses mass and expands in size, while the system evolves to longer
orbital periods.  The minimum orbital period at which mass transfer
can begin is $\sim$ two minutes for a 0.45 $M_{\odot}$ and 0.25 $M_{\odot}$ binary.  Maximum orbital periods for evolved
systems in the Galaxy are estimated to be about 90 minutes; these
correspond to very old systems whose orbits have slowly evolved over a
Hubble time.

Surveys specifically targeted at low-mass WD [such as the ELM survey of WDs: \cite{Brown:2010,Kilic:2011,Brown:2011}] have resulted in
the discovery of numerous DDWD systems with $q<
2/3$, which are expected to get into contact within a Hubble time. Several of these newly discovered low-mass binaries will form stable mass-transferring AM CVn systems, even in
the most pessimistic case of no tidal coupling \citep{Brown:2011}. 

\subsection{The Helium star--white dwarf channel}

In this channel, a WD accretes from an initially nondegenerate He star with mass
0.4--0.6 $M_{\odot}$ before the He has been exhausted in the He-star
core. He stars are produced from stars with masses greater than 2
$M_{\odot}$ that lose their envelopes on the red-giant branch, and
typically have lifetimes comparable to their main-sequence progenitors, allowing for
mass transfer before evolution to the WD stage.     
Mass transfer is stable when the
mass ratio of the donor He star to the accretor WD is less than
approximately 1.2 (\citealt{Tutukov:1989,Ergma:1990}; see also \citealt{NPVY01}).  
Due to mass loss, He burning in the donor's core quenches and at
masses $\sim$ 0.2 $M_{\odot}$
the star becomes semi-degenerate. The minimum orbital period reached by these systems is $\sim$ ten minutes.
   
It remains in question whether 0.1 $M_{\odot}$ or 0.3 $M_{\odot}$ of
He accreted from a He
star will destroy a CO WD by edge-lit detonations (henceforth denoted ELDs; see \citealt{Shen:2010} and
references therein). \cite{NPVY01} distinguish between optimistic and pessimistic assumptions for this channel depending on whether double edge-lit detonations remove CO WD systems.

\subsection{The evolved Cataclysmic Variable channel}

In analogy to traditional CVs, a so-called evolved CV comprises a main-sequence star
which transfers mass to a WD or neutron star \citep{Podsiadlowski:2001}. The difference is that mass
transfer occurs as the donor star evolves off the hydrogen main sequence.
However, it is thought that the evolved-CV channel contributes only
marginally ($< 2 \%$) to the AM CVn population
\citep{Nelemans:2004,Roelofs:2007} for systems with periods less than
1500s. Therefore, we do not consider this channel in this work, which focuses on the
detectability of AM CVn systems by GW interferometers.

\section{Observations of AM CVn and DDWD systems in the Galaxy}
\label{sec:revpop}

We now turn our attention to recent population estimates of AM CVn
systems based on observations.  As mentioned, \cite{Roelofs:2007} estimate the total space density
of AM CVn systems using observations of a source population identified by emission lines in
SDSS. Characterizing selection
effects for the six observed systems in SDSS is challenging.
The orbital periods of these systems are generally shorter than predicted for the majority of AM CVn systems by population estimates, thereby requiring a significant extrapolation to estimate
the space density of the entire population.

Table 1 in \cite{Roelofs:2007} presents the modeled and
inferred AM CVn local space densities for different population-synthesis models. The
inferred densities range from $1.2 \times 10^{-6}$ to $3.4
\times 10^{-6} \, \, \mathrm{pc}^{-3}$, assuming a thin-disk population with
a scale height of 300 pc for most systems.
A comparison between the inferred and modeled densities immediately disfavors
formation models such as the
optimistic DDWD channel, and requires an overall reduction in the 
estimated number of Galactic AM CVn systems by a factor larger than ten.
\cite{Roelofs:2007} hypothesize that lower AM CVn densities arise
predominantly due to the lower efficiency of the major formation channels. 
Later in this paper we consider an alternative possibility, namely that systems observed by \cite{Roelofs:2007} may come from a
distribution with scale height larger than conventional thin disks, with AM CVn
progenitors born early in the history of the Galaxy, before the thin disk was fully developed.

Although the number of known DDWD systems that will merge
within a Hubble time has increased to 24, an approximate factor of five, in the last few years (four eclipsing systems are now
known; the ELM survey has uncovered the eclipsing DDWD binary with the shortest known period of 12.7 minutes), a detailed population
analysis of such systems has yet to be undertaken. More recently, on the basis of the
assumed Galactic-disk model, the ELM survey
predicts $40$ kpc$^{-3}$ of DDWDs where at least one WD has a mass $<$
0.25 $M_{\odot}$, with a factor of two uncertainty in correction factors
\citep{Brown:2011}. This is much
lower than population-synthesis results
(\citealt{Nelemans2001ddwd}). On the other hand, population synthesis results show good agreement with observations made by
the Supernova type Ia Progenitor (SPY) survey. The
SPY radial-velocity survey estimated the local space
density of WDs to be $4.8 \times 10^{-3}$pc$^{-3}$ \citep{Holberg:2008},
with $\sim 100$ new DDWDs found (with orbital periods less than
ten days) from a sample of $1000$. At present, estimating space densities from the ELM
and SPY surveys is very challenging, requiring proper understanding of selection biases. Detailed population comparisons between
the two surveys are not currently available. 

\section{Our analysis}
\label{sec:analysis}

We now describe our analysis: in Sec.\ \ref{sec:scenarios}
we explain our choices for WD binary populations; in Sec.\ \ref{sec:detecting} we discuss
GW emission from binaries; and in Sec.\ \ref{sec:simulations} we describe our simulations of
binary detection and subtraction in detector data. Sec.\ \ref{sec:results} below presents the resulting estimates of the number of detectable systems and of the residual confusion-noise foreground.

\subsection{WD binary catalogs}
\label{sec:scenarios}

In our analysis we considered several different assumptions on AM CVn formation,
generated simulated binary catalogs and detector datasets for each,
and estimated the resulting binary detection and subtraction prospects.
Among the large number of cases that we investigated, here for conciseness we report 
on five scenarios that we believe  are representative of the range of
possibilities given the constraints of current observations. 
Two of the scenarios represent an upper bound (with maximally efficient
formation channels) and conversely, a bound derived assuming inefficient formation channels on predictions from population-synthesis models (cases 1 and 2); two others attempt to reproduce
SDSS-estimated AM CVn densities (cases 3 and 4 below); 
a fifth (case 5) explores a Galactic population with a
larger scale height at times earlier than 7 Gyr ago. As discussed below, the modified--Galactic-disk scenario (case 5) should be consistent with SDSS
observations; however, this agreement is difficult to quantify as the
SDSS-estimated densities given in \cite{Roelofs:2007} are derived
assuming a standard disk model.

We generate DDWD and AM CVn catalogs with the \texttt{SeBa}
population-synthesis code (\citealt{PortegiesZwart:1996,NYP01,Nelemans:2004,Toonen:2011}),
with the following assumptions and parameters: i) the time- and
position-dependent star formation rate (SFR) based on 
\cite{Boissier:1999} (see \citealt{Nelemans:2004}); ii) initial primary masses distributed with power-law
index $-2.5$ (see \citealt{Kroupa:1993}); iii) a flat initial mass-ratio distribution; iv) a flat distribution for the logarithm of the semimajor axis, up to a radius of $10^6
\, R_{\odot}$; v) eccentricities distributed as $P(e)\propto
e^2$; vi) a 50\% binary fraction in the initial population of
main-sequence stars. We expect a factor of two difference in population-synthesis results obtained using different assumptions and initial parameters
than those stated above (\citealt{PortegiesZwart:1996,NYP01,Nelemans:2004}).
Each system is described
by seven parameters: the binary masses $M_1$ and $M_2$, the orbital
period $P_\mathrm{orb}$, the mass-transfer rate $\dot{M}$, the Galactic latitude $l$
and longitude $b$, and the luminosity distance $d$ to the Solar System. 

Following \cite{Nelemans:2004}, we distribute
single and binary stars according to a Galactic model 
that comprises both disk and bulge components.\footnote{The bulge is
  not included in the analysis of \cite{Roelofs:2007}. Furthermore, both
  \cite{Ruiter:2009} and \cite{Yu:2010} also include a halo component;
  however, the impact of halo DDWD systems seems negligible for GW
  interferometers.} The density of systems in the disk is
\begin{equation} 
\label{eq:galdiskmodel}
\rho_\mathrm{disk} (R,Z) = \rho_{\rm BP} (R) \, \mathrm{sech}^2 \biggl(
  \frac{Z}{Z_{\mathrm disk}}
\biggr) \, \mathrm{pc}^{-3},
\end{equation}
where $R$ is the cylindrical radius from the Galactic center; $\rho_{\rm BP}(R)$ results from integrating the time-dependent plane-projected SFR of \cite{Boissier:1999} [see Eqs.\ (1) and (2) of \cite{Nelemans:2004}]; $Z$ is the perpendicular distance from the Galactic plane; and $Z_\mathrm{disk}$ is the disk's scale height.
The density of systems in the spherical bulge is given by
\begin{equation} 
\label{eq:galbulge}
\rho_\mathrm{bulge} \, (r) \propto e^{-(r/r_\mathrm{bulge})^2} \mathrm{pc}^{-3},
\end{equation}
where $r$ is the spherical distance from the Galactic center;
$r_\mathrm{bulge}$ is the characteristic radius of the bulge;
and Eq.\ \eqref{eq:galbulge} is normalized so that there are as many bulge systems as disk systems in the inner 3 kpc of the Galaxy, with a current total mass of $2.6 \times 10^{6} \, M_{\odot}$.

For the first four scenarios we set
$Z_\mathrm{disk} = 300 \, \mathrm{pc}$ and
$r_\mathrm{bulge} = 0.5 \, \mathrm{kpc}$ \citep{Juric:2008,Nelemans:2004,BinneyTremaine:1987,Sofue:2009};
$\rho_\mathrm{BP}(R)$ is fit well by a density $\propto \exp(-R/R_\mathrm{disk})$ with $R_\mathrm{disk} \approx 2.5$ kpc for DDWD systems and $R_\mathrm{disk} \approx 2.2\mbox{--}2.4$ kpc for AM CVn systems.
In the last scenario we explore a modified model of the Galactic disk
with a larger scale height for binaries that are produced at earlier
times ($> 7 $ Gyr ago).

For all but one scenarios we use the DDWD catalog analyzed in
\cite{Nelemans:2004}, with local density $1.2 \times 10^{-5} \,
\mathrm{pc}^{-3}$, and orbital periods as large as 5.5 hrs
($f_\mathrm{GW} = 10^{-4}$ Hz).  Binaries at longer periods are not
typically detectable by GW instruments because detector noise is
higher at low frequencies; they are therefore not considered further
in this analysis. All the catalogs of AM CVn systems, by contrast, show the effects of our different assumptions on AM CVn formation:
\begin{description}
\item[Case 1] upper bound (strong tidal coupling and inefficient ELD events). This catalog includes AM CVn
  systems that have formed via the DDWD channel operating with strong tidal
  coupling and the He
  star--WD channel with few ELD events (see \citealt{NYP01,Nelemans:2004}). Although the resulting AM CVn local density is
  higher ($2.8 \times 10^{-5} \, \mathrm{pc}^{-3}$) than
  SDSS-estimated values, past investigations of Galactic WD binaries as LISA sources have all used variants of this scenario
(see, e.g., \citealt{bibgs:edlund+05,Timpano:2006}). 
Thus, this case provides a meaningful comparison between our study and previously published results, which may vary depending on their idealizations of ``detection,'' ``resolution,'' etc. (see Sec.\ \ref{sec:simulations}).
\item[Case 2] inefficient formation channels (weak tidal coupling and efficient ELD events). This catalog includes AM CVn
  systems that have formed via the DDWD channel operating with minimal tidal
  coupling and the He
  star--WD channel with many ELD events (see \citealt{NYP01,Nelemans:2004}). The AM CVn local density is $6.1 \times 10^{-6}
  \, \mathrm{pc}^{-3}$, a factor of two greater
than SDSS-estimated values \citep{Roelofs:2007b}, and fall within
the factor of two uncertainty of population-synthesis results.
\item[Case 3] mild tidal coupling and no ELD events. This catalog includes AM CVn
  systems that have formed via only the DDWD channel, where mild tidal
  coupling ($\tau_s$ = 2 years) between the WDs allows for stable mass
transfer in up to 10\% of systems compared to the optimistic channel of Case 1.
We assume that the He star--WD channel is suppressed.
[However, observations discussed in \citep{Roelofs:2007b} suggest the presence of
hot donors, which imply either that the He star--WD
channel is active with core He burning having been quenched only recently, or
that following the second common-envelope ejection, the initial DDWD orbital separation is so small that the donor WD cannot cool within the gravitational-radiation timescale.] The AM CVn local density is $1.1 \times 10^{-6} \, \mathrm{pc}^{-3}$.
\item[Case 4] fewer progenitor DDWDs. In this scenario we explain the
SDSS-estimated AM CVn densities by postulating a 1/5 reduction in the
number of their DDWD progenitors. This reduction in DDWD systems is
consistent with existing observations and with the factor of
two uncertainty introduced in population-synthesis models
by our limited understanding of binary evolution and Galactic structure. This catalog also includes AM CVn
systems that have formed both via the DDWD channel with mild tidal
coupling ($\tau_s$ =  1 year) and via the pessimistic He star--WD
channel (see \citealt{NYP01,Nelemans:2004}); their densities are also reduced by 1/5 (across all
periods). The DDWD and AM CVn local densities are $
2.4 \times 10^{-6}$ and
$1.5 \times 10^{-6} \, \mathrm{pc}^{-3}$.

\item[Case 5] modified Galactic-disk model. 
In this scenario we place 
older systems in a thicker disk. With the caveat of small-number statistics,
AM CVn systems identified by emission lines in SDSS
have inferred projected velocities $> 70$ km s$^{-1}$, hinting that AM CVn
systems with longer periods may originate from
a disk with a thick component, or at least from a thin disk with a larger scale
height. Indeed, such a distribution of AM CVns at birth is consistent with our
current (if everchanging) understanding of the spatial and temporal evolution
of thin and thick galaxy disks \citep{Scannapieco:2011,Juric:2008,Schonrich:2009a,Schonrich:2009b,Siebert:2011,Bovy:2011}. 
Observational constraints are unfortunately limited (see \citealt{Napiwotzki:2009} for a discussion on the
thin-disk, thick-disk, and halo populations of single WDs).
As an illustrative example, we altered the disk component of Eq.\ {\ref{eq:galdiskmodel}} to have 
$Z_\mathrm{disk} = 1250 \, \mathrm{pc}$ for systems older than 7 Gyr,
$Z_\mathrm{disk} = 300 \, \mathrm{pc}$ for systems younger than 7 Gyr, and
$r_\mathrm{bulge} = 0.5 \, \mathrm{kpc}$ as for cases 1--4. 
This effectively places all systems older than 7 Gyr in a thick disk,
which includes the great majority of longer-period systems. In
addition, we assumed that AM CVn systems only formed via the DDWD
channel. The DDWD and AM CVn local densities are $1.2 \times 10^{-5}$ and $3.8 \times 10^{-6} \, \mathrm{pc}^{-3}$.
\end{description}
For each scenario, Table~\ref{tab:binpop} shows the numbers of AM
CVn and DDWD systems in the catalogs, as well as their local density
$\rho_0$ (i.e., the Galactic mid-plane density at 8.5 kpc from the
Galactic center). 
\begin{table*}
\caption{Total number and local density of compact-binary systems of different classes (detached, AM CVn from the DDWD channel, AM CVn from the He-star--WD channel) in our simulated catalogs for cases 1--5.\label{tab:binpop}}
\centering
\begin{tabular}{lr@{\;/\;}lr@{\;/\;}lr@{\;/\;}l} \hline \hline
& \multicolumn{6}{c}{number in catalog / local space density [$\mathrm{pc}^{-3}$]} \\
& \multicolumn{2}{c}{DDWD} & \multicolumn{2}{c}{AM CVn (DDWD ch.)} & \multicolumn{2}{c}{AM CVn (He-star--WD ch.)} \\
\hline
case 1 & $26,084,411$ & $1.2 \times 10^{-5}$ & $23,025,764$ & $1.9 \times 10^{-5}$ & $11,197,735$ & $9.0 \times 10^{-6}$ \\
case 2 & $26,084,411$ & $1.2 \times 10^{-5}$ & $261,840$    & $2.1 \times 10^{-7}$ & $6,643,091$  & $5.9 \times 10^{-6}$ \\
case 3 & $26,084,411$ & $1.2 \times 10^{-5}$ & $3,178,553$  & $1.1 \times 10^{-6}$ & none         & 0 \\ 
case 4 & $5,217,866$  & $2.4 \times 10^{-6}$ & $994,194$    & $2.6 \times 10^{-7}$ & $1,328,909$  & $1.2 \times 10^{-6}$ \\
case 5 & $26,084,509$ & $1.2 \times 10^{-5}$ & $23,025,764$ & $3.8 \times 10^{-6}$ & none         & 0 \\
\hline \hline
\end{tabular}
\end{table*}

\begin{figure*}
\includegraphics[width=\textwidth]{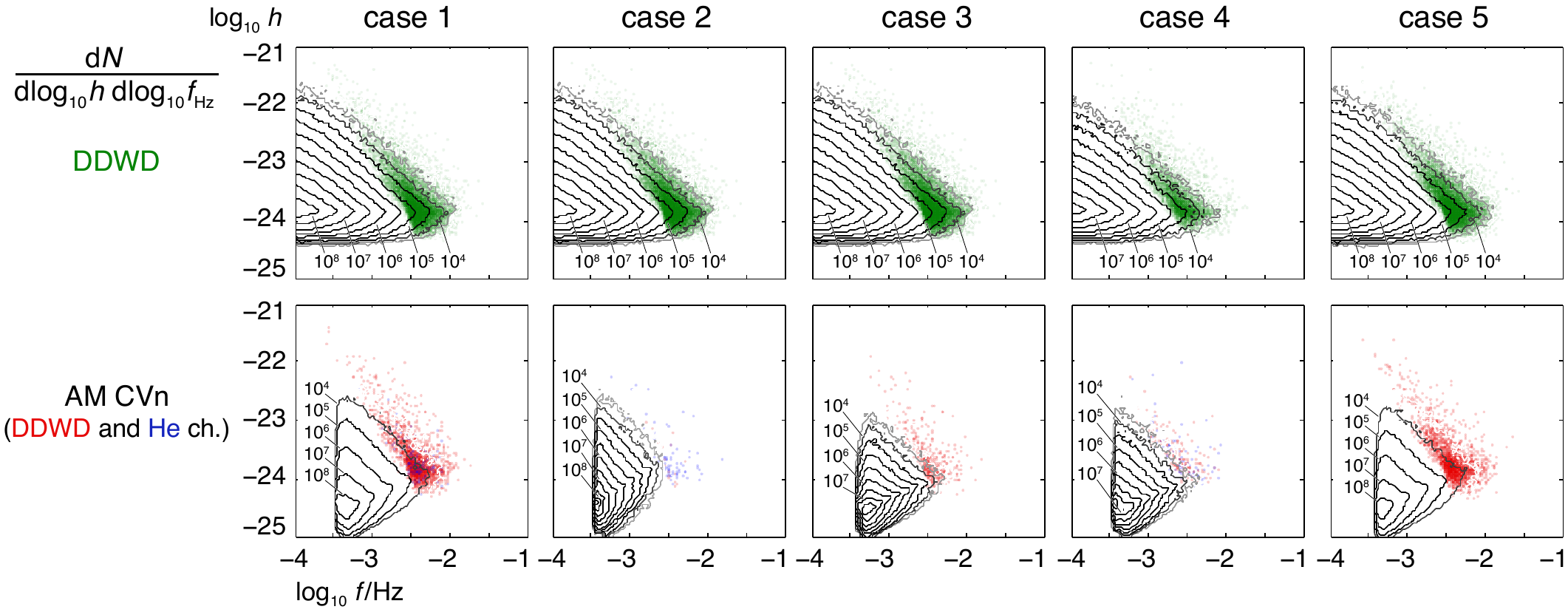}
\caption{Number density as a function of GW-strain amplitude $h$ and
  GW frequency $f$ for DDWD (top row) and AM CVn (bottom row) systems
  in our five scenarios. Green dots denote
  individually detected DDWD systems (for one year of observation with the 5-Mkm detector); red
  dots AM CVn systems from the DDWD channel,
  and blue dots AM CVn systems from the He
  star channel. See Secs.\ \ref{sec:simulations} and
  \ref{sec:results_det} for a discussion of individually resolvable
  sources. The noticeable low-frequency cutoff of AM CVn
  systems is due to the age of the Universe.
  \label{fig:pops}}
\vspace{0.1in}
\end{figure*}

\subsection{Detecting gravitational waves from WD binaries}
\label{sec:detecting}

Compact Galactic binaries in the low-frequency band emit quasi-monochromatic GWs, with frequency drifts of (at most) several frequency bins for year-long observation. 
In this paper, we model compact-binary waveforms by including only the dominant quadrupolar emission \citep{Peters:1963ux}, with instantaneous strain amplitude 
\begin{equation}
h = 10^{-21}
\biggl(\frac{\mathcal{M}_z}{M_{\odot}}    \biggr)^{5/3}
\biggl(\frac{P_{\mathrm{orb}}}{\mathrm{hr}}\biggr)^{-2/3}
\biggl(\frac{d}{\mathrm{kpc}}             \biggr)^{-1},
\label{eq:wave}
\end{equation}
%
where ${\mathcal M_z}= (1+z) \mu^{3/5} M^{2/5}=(1+z){\mathcal M}$ is the redshifted chirp mass
(with $M$ the total mass, $\mu$ the reduced mass, $\mathcal M$ the
chirp mass, and $z$ the redshift), $d$ is the luminosity distance to the source,
and $P_{\mathrm{orb}}$ is the orbital period of the binary.

Figure~\ref{fig:pops} shows the number density as a
function of $h$ [the GW strain of Eq.\ \eqref{eq:wave}], and $f=2/P_{\mathrm{orb}}$ [the GW frequency], for the different DDWD and AM CVn catalogs of each of our scenarios.
The colored dots indicate detected binaries (see Sec.~\ref{sec:simulations}): green
shows DDWDs, red shows AM CVn systems (from the DDWD channel) and blue
shows AM CVn systems (from the He-star--WD channel).

Binaries with orbital periods shorter than approximately 20 minutes
will appear as isolated signals in year-long datasets of space-based
GW observatories, and will be detectable when the signals are
sufficiently strong compared to instrument noise at the same
frequency. By contrast, most binaries with longer orbital periods will not be detected individually, because they will be too dense in frequency space to tell apart. These will produce a noise-like confusion foreground that will affect the detection and parameter estimation of extra-Galactic GW sources such as massive--black-hole binaries. Nevertheless, the brightest binaries will be detected above the confusion foreground.

Looking at Fig.\ \ref{fig:pops}, we see that DDWD systems have larger densities than AM CVn systems, and larger GW strains (because they have greater chirp masses). Thus it is DDWD systems that are principally responsible for the shape and strength of
the GW foreground from WD binaries (see Sec.\ \ref{sec:results_fore}). We note also that all AM CVn populations show an overall low-frequency cutoff at $\log_{10} f = - 3.5$, due to the age
of the Universe.

A simplified (but roughly correct) description of GW data analysis is that GW signals will be detectable when their signal-to-noise ratio, SNR, is greater than a detection threshold $\mathrm{SNR}_\mathrm{thr}$ chosen to yield a minimal number of false alarms. SNR is the ratio of signal strength to rms instrument noise (plus confusion noise, if present) at the same frequency. For LISA-like interferometric detectors, the yearly orbit and rotation of the spacecraft constellation imprint the signals with frequency and amplitude modulations that encode the sky position of the source. Detector response is further affected in frequency-dependent fashion by Time-Delay Interferometry (TDI), the technique used to subtract the otherwise overwhelming laser frequency noise (thus, SNR would be evaluated for the TDI signal with respect to TDI noise). See App.\ \ref{sec:interferometers} for an overview of LISA-like detectors, their response to GWs, and the theory of probabilistic signal detection in noise.

\subsection{Simulations of detection and subtraction}
\label{sec:simulations}

To estimate the number of Galactic-binary systems that we will be able
to detect individually with our reference detectors, as well as the
confusion noise from the residual unsubtracted systems, we performed
simulations of detection and subtraction using a variant of the method
introduced by \cite{Timpano:2006} (see Sec.~\ref{sec:results_det}). To wit:
\begin{enumerate}
\item For each of our scenarios we compute the frequency-domain TDI response for each system in the corresponding catalog, using an optimized variant \citep{Vallisneri:2011} of the Mock LISA Data Challenge fast-binary code \citep{Cornish:2007}; we then superpose the responses to produce a full year-long dataset.
\item We run through the catalog and evaluate the SNR of each
  system (see Sec.\ \ref{sec:interferometers}) with respect to a
  background of instrumental noise plus confusion noise from all
  nearby systems, estimated as a running median of their power
  spectral density. We flag as detected the systems with $\mathrm{SNR} > \mathrm{SNR}_\mathrm{thr}$.
\item We subtract the signals detected in step 2 from the dataset. We then go back to step 2 and iterate, using the resulting lower estimate of confusion noise.
\item The process converges after a few iterations, with the number of detected systems decreasing better than exponentially. We then evaluate and fit the power spectral density from the residual confusion foreground.
\end{enumerate}
This process represents a rather optimistic model of source detection and subtraction that neglects two important effects: source detection will be confused by the presence of nearby signals beyond what is predicted by the increased noise level, and subtraction will be degraded by the imperfect estimation of parameters in noise \citep[see, e.g.,][]{Crowder:2004}.
Nevertheless, we can adjust the level of optimism by varying our assumptions regarding the duration of observation $T_\mathrm{obs}$, the detection threshold $\mathrm{SNR}_\mathrm{thr}$, and the number $N_\mathrm{obs}$ of TDI observables that are available (two, in effect, for the standard LISA configuration with five or six operating links; one for a two-arm configuration).

It turns out that our results for the number of detected sources and for the residual confusion foregrounds are fit remarkably well by power laws of a single \emph{effective-SNR} parameter $\rho_\mathrm{eff} = \mathrm{SNR}_\mathrm{thr} (T_\mathrm{obs}/\mathrm{yr})^{-1/2}(N_\mathrm{obs})^{-1/2}$. As a baseline we take a standard assumption, consistent with prior literature, of one year, one observable (the standard unequal-arm Michelson $X$), and $\mathrm{SNR}_\mathrm{thr} = 5$. A conservative assumption would correspond to taking one year, one observable, and $\mathrm{SNR}_\mathrm{th} = 10$, yielding $\rho_\mathrm{eff} = 10$; an optimistic assumption to five years, two independent observables, and 
$\mathrm{SNR}_\mathrm{th} = 5$, yielding $\rho_\mathrm{eff} = 1.58$. Results are not sensitive to details such as the choice of running-median window, and whether signals are subtracted immediately or after each full catalog run-through.

\section{Results}
\label{sec:results}

\begin{figure*}
\begin{center}
\includegraphics[width=0.9\textwidth]{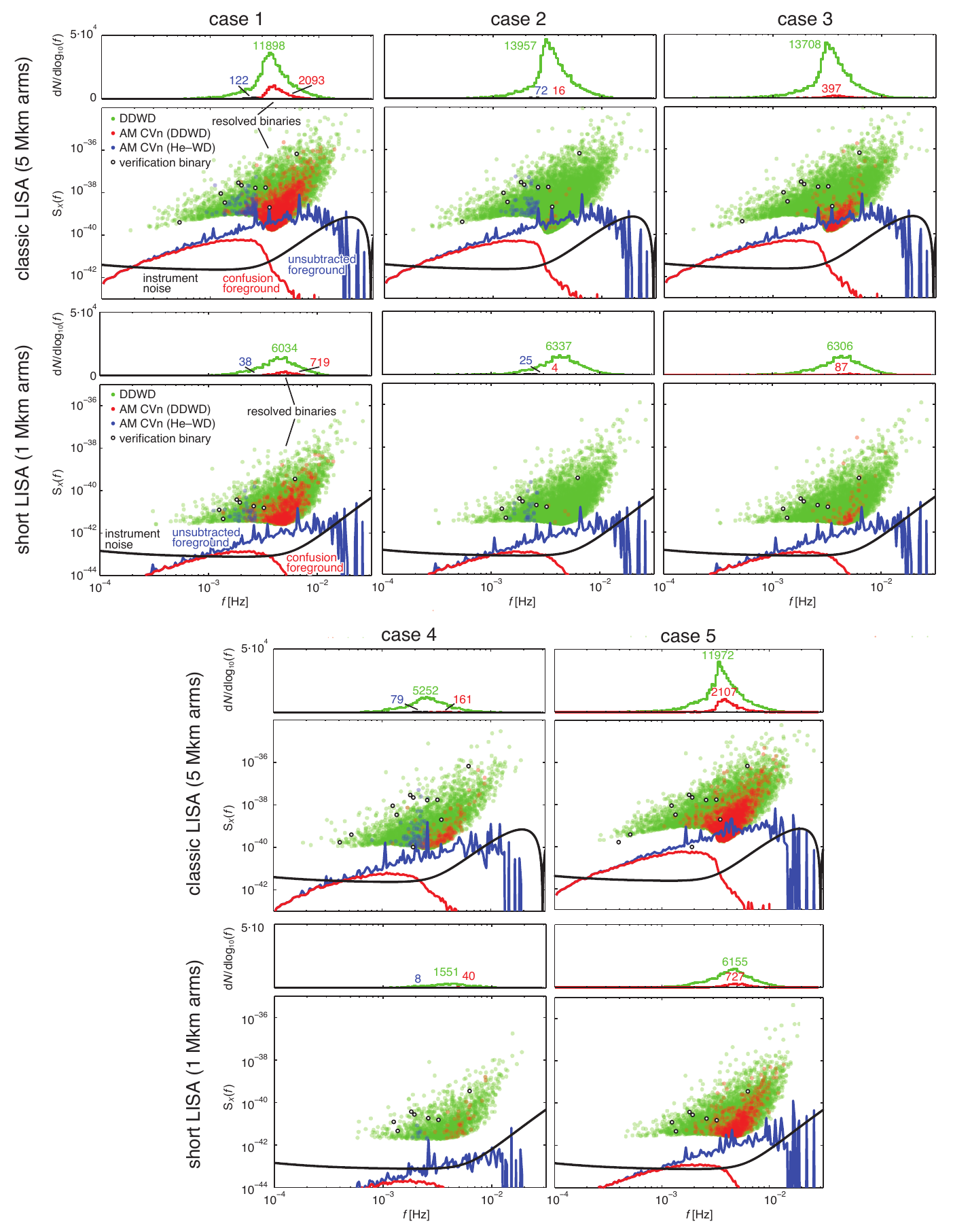}
\end{center}
\caption{Frequency-space density and GW foreground of DDWD and AM CVn systems for
the five scenarios examined in this paper. In each subplot, the bottom panel shows the power spectral density of the unsubtracted (blue) and
partially subtracted (red) foreground, compared to the instrument
noise (black). The open white circles indicate the
frequency and amplitude of the so-called ``verification
binaries'' (see \citealt{Nelemans}). These comprise known DDWD and AM CVn
systems; a handful have known distances from HST FGS parallax
observations, while we assume that the rest are at a distance of 1 Kpc. The
green dots show the individually detectable DDWD
systems; the red dots show the detectable AM CVn systems formed from DDWD progenitors
and the blue dots show the detectable AM CVn systems that arise from the He-star--WD
channel.  The top panel shows histograms of the detected sources
in frequency space. }\label{fig:simulation}
\end{figure*}

In this section we present our results on the impact that different Galactic
populations of DDWD and AM CVn systems have on the number of individually detectable sources (Sec.\ \ref{sec:results_det}) and on the total and confusion GW foregrounds (Sec.\ \ref{sec:results_fore}), for our two representative space-based GW detectors and for each of our five population scenarios.

Figure~\ref{fig:simulation} recapitulates the findings of our simulations:
it shows the frequency-space density of catalog sources; the power spectral density of the unsubtracted (blue) and
subtracted (red) confusion foreground, compared to the instrument
noise (black, all in the bottom panel); and the location and numbers
of detected sources (dots in the bottom panel, histograms in the top panels). The detected sources lie
approximately a factor of $\mathrm{SNR}_\mathrm{th}^2$ above the
composite of instrument noise and subtracted confusion noise.

\subsection{Individually detected sources}
\label{sec:results_det}

Tables \ref{tab:detLISA} (for the 5 Mkm detector) and \ref{tab:detshortLISA} (for the 1 Mkm detector)
show the number of recovered systems of each type in our five scenarios.
The numbers quoted correspond to simulation results for one year, one interferometric observable,
and $\mathrm{SNR}_\mathrm{thr} = 5$; the power laws give approximate fits as functions of
$\rho_\mathrm{eff} = \mathrm{SNR}_\mathrm{thr} (T_\mathrm{obs}/\mathrm{yr})^{-1/2}(N_\mathrm{obs})^{-1/2}$.

Consider first the number of detectable sources as seen by the longer, standard-LISA--like detector.
In all cases except the ``fewer progenitors'' scenario 4, the number of detected DDWDs remains 
comparable to prior optimistic projections ($\sim 12,000\mbox{--}14,000$). Even in case 4, where the DDWD systems
in the catalog are reduced by a factor of five, detections drop only
by half, thanks to a weaker confusion foreground.
By contrast, the number of detected AM CVn varies considerably depending on their density and formation scenarios
($\sim 15\mbox{--}2,000$). In cases 2 and 3, more DDWDs are detectable at relatively high frequencies because of the reduced confusion noise from fewer AM CVn.

Case 1 was chosen to be directly comparable to the standard population
catalogs of earlier works
\citep{NPVY01,NYP01,Nelemans2001ddwd,Nelemans:2004}, but the claimed number of detections for AM CVn systems is significantly lower than
those earlier estimates, derived with a simple resolvability criterion of a single source per frequency bin
with instrument-noise $\mathrm{SNR} > 1$ \citep{NYP01,Nelemans:2004}.
This is unsurprising, given that in our analysis we model detection by identifying and subtracting bright sources iteratively,
and that we require a higher SNR, defined with respect to both
instrument and confusion noise. Furthermore, in this paper we use proper
interferometric observables instead of ``raw'' GW strain. 

Downscaling the detector's armlength to 1 Mkm approximately halves the
number of DDWD systems detected for $\rho_\mathrm{eff} = 5$, except
for case 4, where the decrease is $\sim 3$. The number of AM CVn system detections is likewise reduced by factors $\sim 3\mbox{--}4$. Note however that the $\rho_\mathrm{eff}$ scalings of Tables 
\ref{tab:detLISA} and \ref{tab:detshortLISA} are slightly different.
The ESA eLISA/NGO study \citep{esastudy} finds numbers compatible with ours, once we take into account our noise assumptions for the 1 Mkm detector, which are somewhat more favorable than NGO's.

\newcommand{\rescale}[2]{$#1 \times (\rho_\mathrm{eff}/5)^{-#2}$}
\begin{table*}
\caption{Detectable sources in our simulations (5 Mkm
  detector). Values are shown for $\mathrm{SNR}_\mathrm{thr} = 5$, one
  interferometric observable, and one year of observation, with
  approximate scalings as a function of $\rho_\mathrm{eff} =
  \mathrm{SNR}_\mathrm{thr}
  (T_\mathrm{obs}/\mathrm{yr})^{-1/2}(N_\mathrm{obs})^{-1/2}$. \vspace{0.2in} \label{tab:detLISA}}
\centering
\begin{tabular}{lrrr}
\hline \hline
& \multicolumn{1}{c}{DDWD} & \multicolumn{1}{c}{AM CVn (DDWD)} & \multicolumn{1}{c}{AM CVn (He--WD)} \\
\hline
case 1 &
\rescale{11,898}{1.4} & \rescale{2,093}{1.8} & \rescale{122}{2.8} \\
case 2 &
\rescale{13,957}{1.4} & \rescale{16}{1.7} & \rescale{72}{2.9} \\
case 3 &
\rescale{13,708}{1.4} & \rescale{397}{2.0} & \multicolumn{1}{c}{none} \\
case 4 &
\rescale{ 5,252}{1.2} & \rescale{161}{1.7} & \rescale{79}{1.8} \\
case 5 &
\rescale{ 11,972}{1.4} & \rescale{2,107}{1.7} &  \multicolumn{1}{c}{none} \\
\hline \hline
\end{tabular}
\end{table*}
\begin{table*}
\caption{Detectable sources in our simulations (1 Mkm detector). Values are shown for $\mathrm{SNR}_\mathrm{thr} = 5$, one interferometric observable, and one year of observation, with approximate scalings as a function of $\rho_\mathrm{eff} = \mathrm{SNR}_\mathrm{thr} (T_\mathrm{obs}/\mathrm{yr})^{-1/2}(N_\mathrm{obs})^{-1/2}$. \vspace{0.2in} \label{tab:detshortLISA}}
\centering
\begin{tabular}{lrrr}
\hline \hline
& \multicolumn{1}{c}{DDWD} & \multicolumn{1}{c}{AM CVn (DDWD)} & \multicolumn{1}{c}{AM CVn (He--WD)} \\
\hline
case 1 &
\rescale{6,034}{1.2} & \rescale{719}{1.4} & \rescale{38}{2.6} \\
case 2 &
\rescale{6,337}{1.2} & \rescale{4}{1.4} & \rescale{25}{2.2} \\
case 3 &
\rescale{6,306}{1.2} & \rescale{87}{1.8} & \multicolumn{1}{c}{none} \\
case 4 &
\rescale{1,551}{1.1} & \rescale{40}{1.3} & \rescale{8}{2.2} \\
case 5 &
\rescale{ 6,155}{1.2} & \rescale{727}{1.4} &  \multicolumn{1}{c}{none}  \\
\hline \hline
\end{tabular}
\vspace{18pt}
\end{table*}

\subsection{Unsubtracted and residual compact-binary foregrounds}
\label{sec:results_fore}

\begin{figure*}
\includegraphics[width=\textwidth]{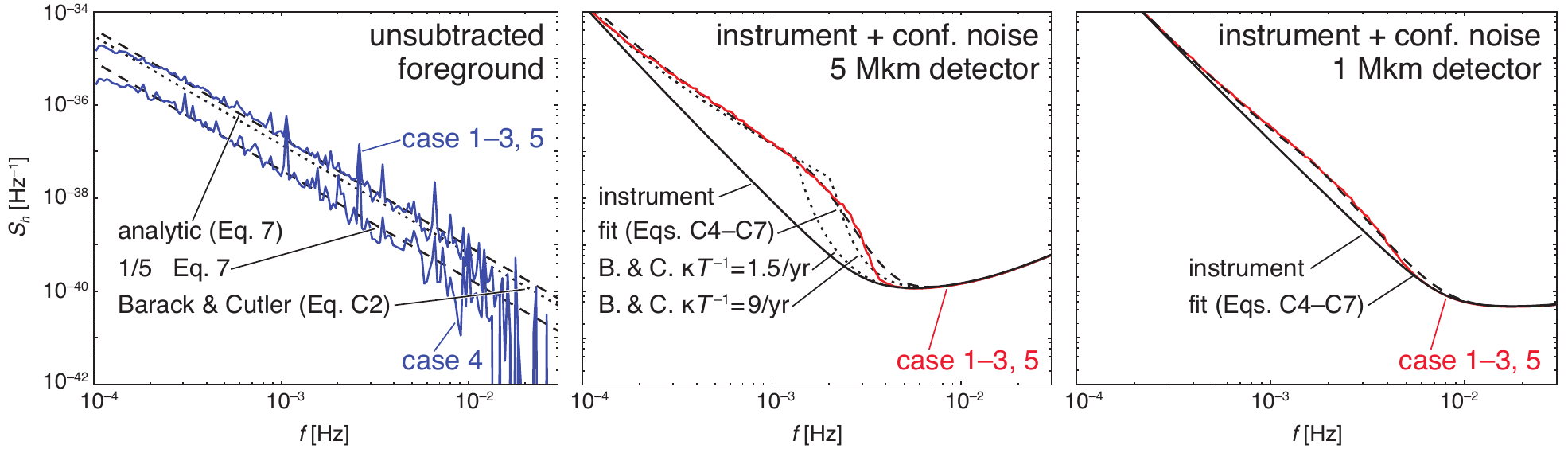}
\caption{\textbf{Left panel}: power spectral density of GW strain for the unsubtracted Galactic foreground, as measured in our simulations for cases 1--3 and 5 (top continuous curve) and case 4 (bottom continuous curve); the foreground is fit well by our Eq.\ 
\eqref{eq:shfshort}, with a 1/5 smaller coefficient for case 4 (dashed curves). For comparison, the dotted curves shows the Barack and Cutler estimate Eq.\  \eqref{eq:BCest}.
At high frequencies, the simulated spectra drop because frequency bins are unevenly occupied, and because of the onset of mass transfer; at low frequency, they drop because some binaries are born at frequencies within the detector's bandwidth, contrary to the assumptions that yield the $f^{-7/3}$ profile.
[These spectra of GW strain do not include the instrument response to GWs; to compare with the TDI $X$ spectra of Fig.\ \ref{fig:simulation}, one would multiply by the frequency-dependent response of Eq.\ \eqref{eq:applyresponse}.]
\textbf{Middle and right panels}: total equivalent GW-strain noise (instrument noise plus residual Galactic-binary confusion noise) for the 5-Mkm and 1-Mkm detectors of our study, as obtained in our simulations for cases 1--3 and 5 and for $\rho_\mathrm{eff} = 5$ (red curves); case 4 is not shown to avoid visual clutter.
The total noise is fit well (dashed curves) by our expressions
Eqs.\  \eqref{eq:theconfusion}--\eqref{eq:thedn} [see also Eqs.\
\eqref{eq:appconf}--\eqref{eq:appconfend}], and less well (dotted
curves) by Barack and Cutler's standard prescription Eqs.\
\eqref{eq:bcone}--\eqref{eq:bcthree}, which improves by setting
$\kappa T^{-1} \simeq 9 / \mathrm{yr}$.
\vspace{0.2in}
\label{fig:fore}}
\end{figure*}

In Appendix \ref{sec:prevwork}, we apply a familiar analytical treatment
of the compact-binary GW foreground to a broad range of Galactic models examined in the literature, and we conclude that the foreground depends strongly on the Galactic binary merger rate and on their characteristic chirp masses, but only weakly on other parameters. Namely,
\begin{eqnarray}
S_h(f) \, & \simeq & \, 1.9 \times 10^{-44} (f/\mathrm{Hz})^{-7/3} \,
\mathrm{Hz}^{-1} \label{eq:shfshort} \\
&\times&
\biggl(\frac{\Deff}{6.4\ {\rm kpc}}\biggr)^{\!-2} \!
\biggl(\frac{\Rtot }{0.015/{\rm yr}}\biggr)
\biggl(\frac{\mathcal{M}_{z, \mathrm{char}}}{0.35\, M_\odot}\biggr)^{5/3},
\nonumber 
\end{eqnarray}
where $\mathcal{D}_\mathrm{char}$ is the characteristic distance from
Earth to Galactic binaries (a weak function of the model);
$\mathcal{R}_\mathrm{gal}$ is the binary merger rate in the Galaxy;
and ${\cal M}_{z, {\mathrm char}}$ is the characteristic chirp mass $\langle {\cal M}_z^{5/3}\rangle^{-5/3}$. The $f^{-7/3}$ slope describes the power emitted at GW frequency $f$ by a \emph{single} binary evolving due to GWs, and it applies to a stationary \emph{ensemble} where new binaries are born with constant rate (equal to $\mathcal{R}_\mathrm{gal}$) at frequencies below the detector bandwidth.
Equation \eqref{eq:shfshort} is normalized to values representative of
our assumptions; as shown in the left panel of Fig.\ \ref{fig:fore}, it is in good agreement with the results of our simulations for cases 1--3 and 5, while the coefficient is reduced by a factor of five for case 4.

The residual confusion foreground consists of the unresolvable DDWD
and AM CVn systems that are left once relatively brighter or isolated
sources have been subtracted iteratively from the original population
catalog. The residual foregrounds found in our simulations are very
similar among cases 1--3 and 5, while they are a factor of five weaker
for case 4. We note that the overall number of DDWD systems in cases 1--3 and
5 differ by a factor of five, thereby confirming earlier work and our
predictions that the DDWD population provides the dominant component to the
strength of the confusion
foregrounds (both unsubtracted and subtracted).  As shown in the middle and right panels of Fig.\ \ref{fig:fore}, for both the 5 Mkm and 1 Mkm detectors, the total equivalent strain noise (instrument noise plus confusion foreground) is fit well by the expression
\begin{equation}
S^\mathrm{total}_h(f) = S_h^\mathrm{inst} + S^\mathrm{conf}_h \times \mathrm{tanh}^\alpha\bigl({\textstyle \frac{1}{2}} \beta \, \mathrm{d}N/\mathrm{d}f \bigr),
\label{eq:theconfusion}
\end{equation}
where $\alpha \simeq 1$, $\beta \simeq 1/\mathrm{yr}$,
\begin{equation}
S_h^\mathrm{conf}(f) = 1.4 \times 10^{-45} \bigl( f / \mathrm{Hz} \bigr)^{-8/3} \, \mathrm{Hz}^{-1},
\end{equation}
and where
\begin{equation}
\frac{\mathrm{d}N}{\mathrm{d}f} = 5 \times 10^{-3} \bigl( f / \mathrm{Hz} \bigr)^{-11/3} \, \mathrm{Hz}^{-1}
\label{eq:thedn}
\end{equation}
is the frequency-domain density of binary systems, which modulates the transition 
from the confused to the resolved regime. Equation \eqref{eq:theconfusion} is formulated in analogy to the standard expression Eq.\ \eqref{eq:bcone} [see \citep{BCu04,BCu04B}], with a different functional form that fits our simulations better. In App.\ \ref{app:subtracted} we give more precise $\alpha$ and $\beta$ fits as functions of $\rho_\mathrm{eff}$, and we provide expressions for the total TDI $X$ noise and for the seasonal modulation of the GW foreground due to detector motion \citep{bibgs:giampieri+polnarev97,2004PhRvD..69l3005S,bibgs:edlund+05}.

\section{Conclusions}
\label{sec:conclusion}

In this work, we have examined the impact that different observationally inferred
populations of Galactic white dwarf binaries have
on the number of individually detectable sources and on the confusion
foreground for future low-frequency GW observations.
Earlier studies of Galactic white dwarf binaries as GW sources
relied primarily on population synthesis, because of the small
number of observed systems, in particular mass transferring ones (i.e., AM
CVn systems). However,
the past several years have seen a flurry of new observations,
quintupling the known detached systems that will merge
within a Hubble time, doubling the known AM CVn systems,
and leading to new, lower estimates for the local densities of the latter.

To understand the consequences of these new estimates on GW detection, 
we have generated a range of plausible populations of white dwarf
binaries, assuming different AM CVn formation scenarios. For example, in one model (case 3) we postulate that fewer AM CVn systems are observed by SDSS than
predicted because the tidal coupling between
the spin of the accretor and the orbit
is not as efficient as expected, so more systems are lost to mergers. We have then investigated the resulting
numbers of individually resolvable GW sources and the magnitude
of the total and residual GW foregrounds.

In four out of five models, we find that the \emph{number of
  detectable detached systems} is $\sim$ 12,000--14,000, for one year
of observation by a LISA-like GW detector with 5 Mkm arms (using an
optimistic detection threshold, but considering a single
interferometric observable). These four models leave unchanged the
number of detached binaries used in earlier estimates, but make
various assumptions about the number of systems lost to mergers or to
edge-lit detonations, or (in case 5) about the spatial distribution of
binaries. In one model (case 4), which instead explains the observed
local AM CVn density with five times fewer detached progenitor
systems, we find that $\sim$ 6,000 of the latter should be
detectable. For the 1-Mkm-arm detector, DDWD detections would be
$\sim$ 6,000 and 1,500 for cases 1-3 and 5, and case 4, respectively.
The \emph{number of detectable AM CVn systems} ranges from several
tens to a few hundred (for cases 2--4), or even as high $\sim$ 2,000
(for cases 1 and 5), again for the 5-Mkm detector. For the 1-Mkm
detector these numbers are smaller by approximately a factor of three to four. See Tables \ref{tab:detLISA} and \ref{tab:detshortLISA} for details.

Now, why is it that even in our upper-bound model (case 1), where the total number of AM CVn systems is comparable to earlier estimates, we project far fewer AM CVn detections than the 10,000 quoted by \cite{NYP01,Nelemans:2004}? Conversely, in four out of five models we project \emph{more} detached-binary detections (12,000--14,000 vs.\ 10,000) than \cite{NYP01}. Such differences arise because we define detection by way of an iterative identify-and-subtract process, where a source's signal-to-noise ratio is evaluated with respect to noise from both the instrument and the partially subtracted foreground. Our approach is similar to that used by \cite{Timpano:2006}. By contrast, according to \cite{NYP01,Nelemans:2004}, a source is detectable if it has a signal-to-noise ratio (relative to instrument noise alone) greater than five, or a signal-to-noise ratio greater than one with no other source in the same frequency bin. Our more stringent definition impacts the number of detected AM CVn systems more than their detached counterparts at the same frequencies, because AM CVn systems have smaller chirp masses, and therefore relatively weaker GW-strain amplitudes.

Indeed, we confirm that the dominant contribution to the \emph{diffuse GW foreground} comes from detached binaries, which have larger chirp masses. In App.\ \ref{sec:prevwork}, we show that the spectrum of the unsubtracted foreground can be derived robustly by simple analytical arguments, and depends mainly on the Galactic merger rate and the characteristic binary chirp mass [see Eq.\ \eqref{eq:shfshort}]. These analytical estimates agree well with our numerical results. In addition, for both the 5 Mkm and 1 Mkm detectors we provide fitting expressions [Eqs.\ \eqref{eq:theconfusion}--\eqref{eq:thedn}] for the residual confusion foreground due to unresolvable sources. Our estimates are comparable to (but measurably different from) the standard formulas discussed by \cite{BCu04B} and used by many authors in the past.

The projected detection numbers and residual confusion foregrounds change significantly if we use different assumptions on the detection threshold, the duration of observation, and the number of available interferometric observables. We have simulated the identify-and-subtract process for several such assumptions, and we find empirically that the detection numbers and confusion-foreground fitting parameters scale as power laws of a single \emph{effective threshold} $\rho_\mathrm{eff} = \mathrm{SNR}_\mathrm{thr} (T_\mathrm{obs}/\mathrm{yr})^{-1/2}(N_\mathrm{obs})^{-1/2}$.

In one of our models (case 5), we have examined (to our knowledge, for the first time)
how GW observations would change if we modify the evolution and structure of the Galactic disk
to match the observed local AM CVn density, while keeping overall binary numbers as in case 1. We find no significant impact on the number of detectable systems, nor on the strength or shape of the confusion
foreground. However, on average the detected systems will have measurably smaller GW amplitudes, because of their larger distances from the Solar system. We conjecture that such distance measurements could be used to infer the structure of the Galactic disk. 

Finally, we expect that the advent of wide-field synoptic and targeted
surveys (such as the  Palomar Transient Factory and the ELM survey respectively), as well as ESA's GAIA mission, will continue to provide new data to this field of research, and will enable more stringent constraints on binary populations. So far, the observed density of detached systems appears to be consistent with theoretical estimates, but a detailed population analysis based on the ELM survey  
has yet to be undertaken. Furthermore, all estimates of AM CVn densities based on surveys are made uncertain by the intricate selection effects. Because of these uncertainties, in this paper we chose to examine a variety of population models that span the range of likely outcomes. Our overall conclusion is that the Galactic-binary science enabled by low-frequency GW detectors is very robust, for a broad range of possible binary populations and detector sensitivities.

\acknowledgements \emph{Acknowledgments.}
For useful discussions and interactions, we are grateful to Lars
Bildsten, Roseanne Di Stefano, Paul Groot, Mukremin Kilic, Shri
Kulkarni, David Levitan, Avi Loeb, Sterl Phinney, Tony Piro, and Ken
Shen. Part of this work was performed by TAP while at the Aspen Center
for Physics, which is supported by NSF Grant \#1066293. GN is
supported by NWO and FOM. MV is grateful for support from the LISA Mission Science Office. Part of this work was performed at the Jet Propulsion Laboratory, California Institute of Technology, under contract with the National Aeronautics and Space Administration. Government sponsorship acknowledged. Copyright 2011.

\appendix

\section{Previous estimates of GW emission from Galactic binaries}
\label{sec:prevwork}
%

\begin{table*} 
\caption{Estimates of GW from Populations of Compact Binaries\label{tab:comparison}}
\centering
\begin{tabular}{l c l l}  
\hline\hline
reference&model&source classes & input population\\[0.5ex]
\hline \hline            
this work, \cite{Nelemans:2004} & RS + CB& DDWD + AMCVn&SeBa$^{1}$\\  
\hline
\cite{NYP01}&RS + CB&DDWD + AMCVn &SeBa$^{1}$\\
\cite{bibgs:edlund+05}&RS + CB& DDWD + AMCVn &SeBa$^{1}$\\
\hline\hline
\cite{Timpano:2006} &RS + CB& DDWD&SeBa$^{1}$, \cite{Webbink:1984}\\
\hline
\cite{bibgs:seto02}&RS + CB& DDWD &\cite{Webbink:1984}\\
\hline
\cite{bibgs:bender+hils97}&RS + CB&DDWD&\cite{Webbink:1984}\\
\cite{bibgs:hils+90}&CB&DDWD&\cite{Webbink:1984}\\
\hline\hline
\cite{HB00}&CB&AMCVn&\cite{Iben:1984,bibgs:tutukov+yungelson96}\\
\cite{bibgs:hils98}&RS + CB&AMCVn&\cite{Iben:1984,bibgs:tutukov+yungelson96}\\
\hline\hline
\cite{Yu:2010} &RS + CB& DDWD (+ halo)&BSE$^{2}$ + \cite{bibgs:han98}\\
\hline
\cite{bibgs:liu09}; \cite{bibgs:liu+10a}&RS
+ CB&DDWD&BSE$^{2}$ + \cite{bibgs:han98}\\
\hline
\cite{bibgs:farmer+phinney03}&CB&extragalactic DDWD&BSE$^{2}$\\
&&+ AM CVn&\\
\hline \cite{bibgs:webbink+han98}&CB&DDWD&\cite{bibgs:han98}\\ [0.5ex]
\hline\hline
\cite{Ruiter:2010}&RS + CB&DDWD + AMCVn&StarTrack, \cite{bibgs:belczynski+02,bibgs:belczynski+08}\\
\cite{Ruiter:2009}&&(halo)&\\
\hline\hline\cite{bibgs:benacquista+04}&RB + CB&DDWD (+ globular)&Specified in paper\\
\hline\cite{bibgs:postnov+prokhorov98}&CB&DDWD&See \cite{bibgs:lipunov+96}\\
\cite{bibgs:lipunov+95}&CB&DDWD (+ halo)&specified in paper\\
\cite{bibgs:lipunov+87a}&CB&compact binaries&specified in paper\\
\cite{bibgs:lipunov+87b}&&&\\
\hline\cite{bibgs:giampieri+polnarev97}&CB&generic&generic\\
\hline
\cite{EIS87}&CB&DDWD&specified in paper\\
\hline\hline
\multicolumn{4}{ l }{{Notes: RS = resolved sources; CB = confusion background.}}\\
\multicolumn{4}{ l }{{$^{1}$SeBa: \cite{PortegiesZwart:1996,bibgs:zwart+98,Nelemans2001ddwd,Toonen:2011}.}}\\
\multicolumn{4}{ l }{{$^{2}$BSE:
    \cite{bibgs:hurley+00,bibgs:hurley+02}.}}\\
\hline \hline 

\end{tabular}
\vspace{0.2in}
\end{table*} 
\renewcommand{\arraystretch}{1.5}
\begin{table*}
\caption{Estimates of GW from Populations of Compact Binaries -- Spatial Distributions\label{tab:spatial}} 
\centering      
\begin{tabular}{lclllll}  
\hline\hline                
reference&model&$Z_\mathrm{disk}$&$R_\mathrm{disk}$&$\Deff$ (disk)&$r_\mathrm{bulge}$\\
&&(kpc)&(kpc)&(kpc)&(kpc)\\[0.5ex] 
\hline \hline
this work, \cite{Nelemans:2004} & see Sec.\ \ref{sec:scenarios} &0.3&2.5$^{1}$&6.4$^{2}$&0.5\\  
this work (thick disk) & &1.25&2.5$^{1}$&6.8$^{2}$&0.5\\
\hline
\cite{NYP01}&$\mathrm{e}^{-R/R_\mathrm{disk}}\,\mathrm{sech}^2(Z/Z_\mathrm{disk})$&0.2&2.5&5.3&--\\
\cite{bibgs:edlund+05}&&&&&&\\
\hline\hline
\cite{Timpano:2006} & \multicolumn{6}{c}{see \cite{HB00} and \cite{Nelemans:2004}} \\
\hline
\cite{bibgs:seto02}&$\mathrm{e}^{-R/R_\mathrm{disk}}  \mathrm{e}^{-Z/Z_\mathrm{disk}}$&0.09&3.5&4.8&--\\
\hline
\cite{bibgs:bender+hils97}&$\mathrm{e}^{-R/R_\mathrm{disk}}  \mathrm{e}^{-Z/Z_\mathrm{disk}}$&0.09&3.5&4.8&--\\
\cite{bibgs:hils+90}&$\mathrm{e}^{-R/R_\mathrm{disk}}  \mathrm{e}^{-Z/Z_\mathrm{disk}}$&0.09&3.5&4.8&--\\
\hline\hline
\cite{HB00}&ns&ns&ns&--&ns\\
\cite{bibgs:hils98}&ns&ns&ns&--&ns\\
\hline\hline
\cite{Yu:2010} &$\mathrm{e}^{-R/R_\mathrm{disk}}\,\mathrm{sech}^2(Z/Z_\mathrm{disk})$&0.352&2.5&5.6&0.5\\
&$\mathrm{e}^{-R/R_\mathrm{disk}}  \mathrm{e}^{-Z/Z_\mathrm{disk}}$&1.158&2.5&6.5&--\\
\hline
\cite{bibgs:liu+10a}; \cite{bibgs:liu09}&$\mathrm{e}^{-R/R_\mathrm{disk}}  \mathrm{e}^{-Z/Z_\mathrm{disk}}$&0.09&2.5&5.0&--\\
\hline
\cite{bibgs:farmer+phinney03}&--&--&--&--&--\\
\hline
\cite{bibgs:webbink+han98}&$\mathrm{e}^{-R/R_\mathrm{disk}}  \mathrm{e}^{-Z/Z_\mathrm{disk}}$&0.09&4.0&4.7&--\\ [0.5ex]
\hline\hline
\cite{Ruiter:2010}&$\mathrm{e}^{-R/R_\mathrm{disk}}  \mathrm{e}^{-Z/Z_\mathrm{disk}}$&0.2&2.5&5.4&0.5\\
\cite{Ruiter:2009}&(halo)&&&\\
\hline\hline
\cite{bibgs:benacquista+04}&$\mathrm{e}^{-R/R_\mathrm{disk}}\,\mathrm{sech}^2(Z/Z_\mathrm{disk}) $&0.2&2.5&5.3&--\\
\hline
\cite{bibgs:postnov+prokhorov98}&$\mathrm{e}^{-R/R_\mathrm{disk}}  \mathrm{e}^{-(Z/Z_\mathrm{disk})^2}$&4.2&5&7.2&--\\
\cite{bibgs:lipunov+95}&$\mathrm{e}^{-R/R_\mathrm{disk}}  \mathrm{e}^{-Z/Z_\mathrm{disk}}$&0.46&4.6&5.7&see paper\\
\cite{bibgs:lipunov+87a}&$\mathrm{e}^{-R/R_\mathrm{disk}}  \mathrm{e}^{-(Z/Z_\mathrm{disk})^2}$&4.2&10&9.6&--\\
\cite{bibgs:lipunov+87b}&&&&&&\\
\hline
\cite{bibgs:giampieri+polnarev97}&$\mathrm{e}^{-R/R_\mathrm{disk}}  \mathrm{e}^{-(Z/Z_\mathrm{disk})^2}$&0.1&3.5&4.5&ns\\
\hline
\cite{EIS87}&constant density&0.2&10&&--\\
\hline\hline
\multicolumn{7}{l}{{Notes: ns = not specified. Distance to the
    Galactic Center is assumed to be 8.5 kpc for all models.  }}\\
\multicolumn{7}{l}{{Exponential squared bulge unless otherwise
    specified: $\mathrm{e}^{-(r/r_{\rm bulge})^2}$ with $\Deff({\rm bulge}) = 8.4 \, \mathrm{kpc}$.}}\\
\multicolumn{7}{l}{{$^{1}$Numerical fit, DDWD systems only.}}\\
\multicolumn{7}{l}{{$^{2}$$\Deff$ computed for disk + bulge for this case.}}\\
\hline \hline
\end{tabular} 
\vspace{0.2in}
\end{table*}

More than two dozen papers have made estimates of the GW signal from
individually detectable Galactic compact binaries and from the
continuum of unresolvable sources at low frequencies (variously called
``confusion background'' or ``Galactic foreground'') for a LISA-like
GW detector with 5-Mkm arms. An earlier informative review of the
literature is given by \cite{Ruiter:2010}. We note that for a 1-Mkm
detector the foreground decreases substantially, to a level comparable
to instrument noise (see Fig. \ref{fig:fore} in Section \ref{sec:results_fore}). 

Table \ref{tab:comparison} compares a subset (large, but possibly not complete) of relevant published estimates, grouped by population model. For each entry, we show whether the estimates include resolved sources (RS), the continuum background (CB), or both; we show also which classes of binaries are included.
In Table \ref{tab:spatial} we provide relevant scale lengths
for the Galactic disk and bulge distributions, where available;
these are indicative of the range of assumed Galactic models, but they do not provide a complete picture since different density laws also appear.
We do not give quantitative information on possible halo components, which are also frequently used.
The major differences between the various estimates arise from two sets of assumptions:
\begin{enumerate}
\item Assumptions about the relative numbers of systems in terms of total mass, mass ratio, and separation. These arise from different treatments of compact-binary progenitor evolution and of the stability of mass-transferring systems.
Approaches include both analytic estimates and Monte Carlo population-synthesis calculations.
\item Assumptions about Galactic structure and population space densities. These can affect estimates in two ways: first, the magnitude of GW signals will depend on the scale size assumed for the Galactic population; second, the ``calibration'' of 
the local number density of sources using observations of a particular source type will depend sensitively on the assumptions about Galactic structure.
\end{enumerate}

We now present a semi-analytical derivation of the amplitude and shape of the GW continuum. Similar models have been used extensively in GW data analysis, see for example works by \cite{bibgs:hughes02,BCu04,BCu04B}.
Our results show that the magnitude and spectral shape of the continuum are a robust function of the merger rate and characteristic chirp mass of Galactic binary systems, and depend only weakly on Galactic structure.

We follow the arguments presented in \cite{EIS87} and \cite{bibgs:giampieri+polnarev97}.
The GW luminosity of a circular binary and the change in its orbital energy as a function of GW frequency $f = 2 f_\mathrm{orb}$ are given by \citep{Peters:1963ux}
\begin{equation}
{dE \over dt} = {32 \over 5} \pi^{10/3} {G^{7/3} \over c^5} {{\cal M}_z}^{10/3} f^{10/3} , \quad
{dE \over df} = {1 \over 3} \pi^{2/3} G^{2/3} {{\cal M}_z}^{5/3} f^{-1/3};
\end{equation}
where $E$ is the energy of the system and $f$ is the GW frequency.
Combining these expressions yields the frequency evolution of the system,
\begin{equation}
{df \over dt} = {96 \over 5}  \pi^{8/3} {G^{5/3} \over c^5} {{\cal M}_z}^{5/3}  f^{11/3}.
\end{equation} 
If $\Rbirth$ is the birth rate per unit frequency and volume of systems with $f_\mathrm{orb}<f/2$, in a \emph{stationary regime} the density of sources must be
\begin{equation}
\label{eq:dNdf}
{d^2 N \over df d {\bf r}} = {\Rbirth \over {df/dt}} 
= {5 \over 96} \pi^{-8/3} {c^5 \over G^{5/3}}  {{\cal M}_z}^{-5/3}\
\Rbirth  f^{-11/3} .
\end{equation}
By continuity, $\Rbirth$ must also be equal to the merger rate, $\Rmergef$, of systems which merge at $f_\mathrm{orb} > f/2$. 
DDWD systems, which generate most of the GW foreground, merge at $f > 10^{-2}$ Hz, and most of them are born below a few $10^{-4}$ Hz (although estimates vary). Thus, if we restrict Eq.\ \eqref{eq:dNdf} between those two frequencies, we may just set $\Rbirth = \Rmrg$, defined as the total merger rate above $10^{-2}$ Hz.

Now, the direction-averaged GW strain from a circular binary is given by
\begin{equation}
\label{eq:hsq}
h^2 = \langle h_+^2 +  h_\times^2 \rangle = {1 \over
  \pi^2} {1 \over d^2}  {G \over c^3} f^{-2} {dE \over dt} 
  = {32 \over 5} \pi^{4/3} {G^{10/3} \over c^8} {{\cal M}_z}^{10/3} f^{4/3}
\end{equation}
%
see for instance \citep{bibgs:giampieri+polnarev97}, 
where $d$ is the luminosity distance to the binary; the corresponding detector response is
\begin{equation}
\label{eq:hfsq}
h_I^2= \langle h_+^2 F_+^2 + h_\times^2 F_\times^2 \rangle \equiv h^2 \Pol,
\end{equation}
where $F_+$ and $F_\times$ are the antenna-pattern functions (see, e.g., \citealt{Maggiore:1900zz}) and $\Pol$ is the polarization- and inclination-averaged antenna pattern $(\langle F_+^2 \rangle + \langle F_\times^2 \rangle)/2$ in the direction ${\bf \hat r}$.

We combine Eqs. \ref{eq:dNdf}, \ref{eq:hsq}, and \ref{eq:hfsq} to obtain the one-sided detector response density $S_{h_I}$. The yearly orbital motion of LISA-like detectors modulates the amplitude of signals to individual sources; this effect averages out for isotropic source distributions, but not for the anisotropic Galactic-binary foreground \citep{bibgs:giampieri+polnarev97,bibgs:edlund+05}.
We therefore average over a year (while assuming for now that all binaries have the same chirp mass $\Mchirp_z$), to obtain
\begin{eqnarray}
S_{h_I}(f) &=& \biggl\langle \int h^2(d({\bf r}), f)\ \Pol\  {d^2 N \over df d {\bf
      r}}\  d{\bf r} \biggr\rangle_t \nonumber \\
&= & {1 \over 3} \pi^{-4/3} {G^{5/3} \over c^3}   {{\cal M}_z}^{5/3}
f^{-7/3}   \!\! \int   {{\langle \Pol\rangle_t} \Rmrg \over d({\bf r})^2}
d{\bf r} \nonumber \\
& \equiv & {1 \over 3} \pi^{-4/3} {G^{5/3} \over c^3}   {{\cal M}_z}^{5/3} f^{-7/3}\ \Pfac \langle \mathcal{R}_\mathrm{mrg} / d^2 \rangle_{\bf r}. \label{shi}
\end{eqnarray}

Of the two quantities defined in the last line of Eq.\ \eqref{shi}, $\Pfac$ accounts for the time-, merger-rate- and direction-averaged instrument response, and it depends on the orbit and geometry of the detector. For an isotropic distribution of sources and a triangular interferometer with $60^\circ$ angles, $\Pol = 1/5 \times \sin^2(60^\circ) = 3/20$, which is often taken as the average isotropic response for LISA-like detectors (\citealt{bibgs:giampieri+polnarev97}). The factor 1/5 is the average of $(\langle F_+^2 \rangle + \langle F_\times^2 \rangle)/2$ over source locations and polarizations (again, e.g., in \citealt{Maggiore:1900zz}). Again for a LISA-like detector, but for typical disk populations, $\Pol$ is found empirically to vary by $\sim 10\%$ from 3/20.

The quantity $\langle \mathcal{R}_\mathrm{mrg}/{d^2} \rangle_{\bf r}$ is the
merger-rate--averaged inverse squared distance. For a Galactic disk, it depends on the scale height and radial scale length, and of course on the position of the detector within the disk. 
A typical disk density distribution considered in many papers is the squared--hyperbolic-secant disk model,
\begin{equation}
\Rmrg = \frac{\Rtot}{V_{\rm gal}} \, \mathrm{e}^{-R/R_\mathrm{disk}} {\rm sech}^2(Z/Z_\mathrm{disk}),
\end{equation}
where $\Rtot$ is the total merger rate in the Galaxy and $V_{\rm gal}$ is the volume of the Galaxy. We find empirically that $\langle \mathcal{R}_\mathrm{mrg}/{d^2} \rangle_{\bf r}$ varies slowly for $2.5\  {\rm kpc} \le R_\mathrm{disk} \le 10\ {\rm kpc}$ and $0.2\  {\rm kpc} \le Z_\mathrm{disk} \le1\  {\rm kpc}$.
Table \ref{tab:spatial} summarizes the wide range of Galactic models
used in the literature. For most disk models the characteristic
distance to Galactic sources $\Deff \equiv (V_{\rm
  gal}/\,\Rtot)^{-1/2}    \langle \mathcal{R}_\mathrm{mrg} / d^2
\rangle^{-1/2}$ is approximately 5 to 6 kpc, and $\Deff + R_\mathrm{disk}$ is close to 8.5 kpc, the assumed distance to the Galactic center.

Inserting numerical values appropriate for our simulations (cases 1--3) in Eq.\ \eqref{shi} yields
\begin{equation}
S_{h_I}(f) \, \simeq \, 1.9 \times 10^{-44} (f/\mathrm{Hz})^{-7/3} \, \mathrm{Hz}^{-1} \times \Pfac 
\times
\biggl(\frac{\Deff}{6.4\ {\rm kpc}}\biggr)^{\!-2} \!
\biggl(\frac{\Rtot}{0.015/{\rm yr}}\biggr)
\biggl(\frac{\mathcal{M}_{z, \mathrm{char}}}{0.35\, M_\odot}\biggr)^{5/3},
\label{eq:Shf}
\end{equation}
where ${\cal M}_{z, {\rm char}}$ is the characteristic chirp mass $\langle {\cal M}_z^{5/3}\rangle^{-5/3}$.
The power spectral density $\Shf$ of the GW strain is obtained by dropping the 
factor $\Pfac$ (which is close to 3/20 for LISA-like detector).
Equation \eqref{eq:Shf} is in good agreement with the results of our simulations. Note however that the spectra of Fig.\ \ref{fig:simulation} are plotted for the TDI observable $X$ (rather than for strain $h$ or strain response $h_I$), so they must be compared with
\begin{equation}
S_X(f) = \frac{16 \, x^2 \sin^2 x}{1 + 0.6 \, x^2} S_{h_I}(f),
\end{equation}
where $x = 2 \pi f L / c$, and $L$ is the detector armlength.

Our analysis indicates that $S_{h_I}(f)$ and $\Shf$ are relatively
insensitive to the structure of the Galactic disk, consistent with the
results of our simulations, and instead depend strongly on the merger
rate $\mathcal{R}_\mathrm{gal}$ and the characteristic chirp mass
$\mathcal{M}_{z, \rm{char}}$. By constrast, we expect that the local population density of sources, as well as the number of resolvable source as a function of their GW strain, will depend roughly linearly on the disk's scale height.

The transition between the confused and resolved detection regimes is characterized by
$dN/df$, which is determined by the evolution of the chirp frequency and by the absolute merger-rate density of sources. Integrating Eq.\ \eqref{eq:dNdf} over $\mathbf{r}$ we find
\begin{equation}
\label{eqn:dNdf}
\frac{dN}{df} =
5 \times 10^{-3} \, (f/\mathrm{Hz})^{-11/3} \,{\rm Hz}^{-1} 
\times
\left({\Rtot \over {0.015/{\rm yr}}}\right)
\left({M_{z, {\rm char}} }\over {0.35\, M_\odot}\right)^{-5/3},
\end{equation}
consistent with our simulations.

This simple analytical derivation of the GW continuum shows that the
major differences between the various estimates listed in Table
\ref{tab:comparison} arise primarily from two sources: the total
merger rate of systems $\Rtot$ and the composition of the source
population, leading to a different characteristic chirp mass $M_{z, \,
  {\rm char}}$. $\Deff$ is also relevant, but less important.
(Remember however that our derivation is only predictive for frequencies high enough that most binaries form at longer periods---for LISA-like detectors, this is true above $10^{-4}$ Hz. For lower frequencies, the shape of the continuum depends on the detailed accounting of birth rate vs.\ period.)


\section{Low-frequency GW interferometers and their response to binary waves}
\label{sec:interferometers}

LISA \citep{lisasciencecase,Jennrich:2009p1398}, the space-based GW observatory studied jointly by NASA and ESA until early 2011, comprises three spacecraft orbiting the Sun in a quasi-equilateral configuration with armlengths $\sim 5 \times 10^6$ km. LISA measures GW strain by monitoring the oscillating distance between freely-falling test masses, using heterodyne laser interferometry. The spacecraft hover around the test masses to protect them from external perturbations, performing slight orbital corrections with extremely precise micro-Newton thrusters.

Although LISA is sometimes described as a large Michelson interferometer in space, the lasers exchanged by spacecraft are neither split nor reflected. Rather, two lasers on each spacecraft are used to establish six one-way links; each link provides a test-mass--referenced interferometric measurement $y_{ij}$ of the phase (or equivalently, the frequency) of the incoming laser, compared to the local, outgoing beam. In terms of fractional laser frequency, the effect of a plane GW on the links is given by the two-pulse response \citep{1975GReGr...6..439E}
\begin{equation}
y_{ij}(t) = \frac{1}{2} \frac{\hat{n}_{ij} \cdot
\bigl[\mathsf{h}(p_i,t-L_{ij}) -
\mathsf{h}(p_j,t)\bigr] \cdot  \hat{n}_{ij}}{1 - \hat{n}_{ij} \cdot \hat{k}},
\end{equation}
where $\hat{n}^i_{ij}$ is the unit vector from spacecraft $i$ to $j$; $\hat{k}$ the GW propagation vector; $\mathsf{h}(p_k,t)$ the TT-gauge GW strain (see, e.g., \citealt{Maggiore:1900zz}) at the spacecraft position $p_k$ and at time $t$; and $L_{ij}$ the photon time of flight along the arm. Thus, a GW pulse is registered twice in $y_{ij}$: once when it impinges on spacecraft $j$, and a time $L_{ij}$ after it has impinged on spacecraft $i$. In the limit of long GW wavelengths $\lambda \gg L$, the $y_{ij}$ response to plane GWs is therefore proportional to L.

Each $y_{ij}$ measurement is also affected by \emph{displacement} noise from residual test-mass accelerations, \emph{position} noise from optical-path and shot noise in the heterodyne measurements, and, most importantly, by oscillations in the central laser frequency, which are many orders of magnitude stronger than expected GW imprints, and which appear as $y_{ij} = C_{ji}(t) - C_{ij}(t - L_{ij})$, with $C_{ij}$ the frequency noises for the six lasers.
Laser frequency noise is removed by combining multiple one-way measurements with appropriate time delays so that each laser appears in canceling pairs, as in the
\emph{unequal-arm Michelson} observable
\begin{equation}
\begin{aligned}
X(t) = & \bigl[ y_{31}(t) + y_{13}(t - L_{31}) + y_{21}(t - L_{31} - L_{13})
+ y_{12}(t - L_{31} - L_{13} - L_{21}) \bigr] \\
- & \bigl[ y_{21}(t) + y_{12}(t - L_{21}) + y_{31}(t - L_{21} - L_{12})
+ y_{13}(t - L_{21} - L_{12} - L_{31})
\bigr].
\label{eq:unequalmich}
\end{aligned}
\end{equation}
This noise-suppression technique, specific to LISA-like interferometers, is known as \emph{time-delay interferometry} (TDI; see, e.g., \citealt{2005PhRvD..72d2003V} for a review). TDI observables can be seen as \emph{synthesized} interferometers, since their $y_{ij}$ components trace the paths (and reproduce the phase delays) of light across recognizable interferometric configurations (hence $X$'s ``Michelson'' designation). Many different observables are possible with LISA's six laser links, but all can be reconstructed from a set of three observables, such as the unequal-arm Michelson $X$, $Y$, and $Z$ centered on the three spacecraft \citep{2008CQGra..25f5005V}.

In this paper, we assume an rms displacement noise
\begin{equation}
S^{1/2}_\mathrm{pm}(f) = 3 \times 10^{-15} \, \mathrm{m} \, \mathrm{s}^{-2} \mathrm{Hz}^{-1/2} \times \sqrt{1 + \Bigl(\frac{f}{10^{-4} \,\mathrm{Hz}}\Bigr)^{-1}} \times \sqrt{1 + \Bigl(\frac{f}{0.008 \,\mathrm{Hz}}\Bigr)^{4}},
\end{equation}
for each of the proof masses, and a total rms position noise
\begin{equation}
S^{1/2}_\mathrm{op}(f) = 18 \times 10^{-12} \, \mathrm{m}\, \mathrm{Hz}^{-1/2} \times \sqrt{1 + \Bigl(\frac{f}{0.002 \, \mathrm{Hz}}\Bigr)^{-4}}
\end{equation}
for each $y_{ij}$ measurement.

The GW mission currently under study as an ESA-led project \citep{esastudy} would decrease cost primarily by reducing spacecraft and propellant mass, allowing for launch on smaller rockets than envisaged for LISA. Propellant is saved by placing the spacecraft closer together and closer to Earth. At the low-frequency end, the shorter armlengths reduce the response to GWs proportionally; for the same displacement noise, sensitivity also decreases by the same ratio. At the high-frequency end, the laser power available for position measurement increases as $L^{-2}$, since beams are broadly defocused at millions of kms, improving shot noise (but not other optical noises) by an rms factor $L^{-1}$. As a consequence, the sweet spot of the LISA sensitivity shifts to higher frequencies.

Spacecraft mass may also be saved by adopting a two-arm ``mother--daughters'' configuration with only four laser links, two each between the middle ``mother'' spacecraft and one of the ``daughters''. With four links, only one independent TDI observable can be formed, with a loss in sensitivity averaging $\sqrt{2}$ at compact-binary frequencies. Because no trade studies have yet been performed to optimize the payload for the new orbits, in this paper we consider a ``short LISA'' configuration with aggressively reduced $1\times10^6$-km armlengths, and with the LISA displacement and position noises (except for the power rescaling of the position-noise component, which we assume as five times smaller in rms).
The actual sensitivity of a cost-constrained space-based GW observatory is likely to fall somewhere between this short LISA and the classic configuration.

Searches for GWs of known shape proceed by correlating detector data with theoretical waveform \emph{templates} computed for the expected range of source parameters. The statistical confidence of detection for a GW signal $h$ immersed in Gaussian additive noise $n$ is described by the \emph{signal-to-noise ratio} (SNR) of the data $s = h + n$ after filtering by the best-fitting \emph{normalized} template $\hat{t}$. For a single TDI observable $X$, this is given by \citep{1994PhRvD..49.2658C}
\begin{equation}
\rho(s;\hat{t}) =
4 \, \mathrm{Re} \int_0^{\infty} \frac{\tilde{X}_s(f)^* \tilde{X}_{\hat{t}}(f)}{S_X(f)} \,\mathrm{d}f,
\end{equation}
where $\tilde{X}_s(f)$ and $\tilde{X}_{\hat{t}}(f)$ are the Fourier transforms of the TDI data and of the TDI response to waveform $\hat{t}$ respectively; ``${}^*$'' denotes the complex conjugate; and $S_X(f)$ is the noise power spectral density of TDI observable $X$. Normalization amounts to requiring $\rho(\hat{t};\hat{t}) = 1$.

With this noise weighting, the false-alarm probability of SNR exceeding a chosen detection threshold $\rho_\mathrm{thr}$ is proportional to $\exp(-\rho_\mathrm{thr}^2/2)$. Because the many templates used in typical searches amount to repeated independent trials, $\rho_\mathrm{thr}$ is routinely set to relatively large values (e.g., 5--10) to yield one false alarm per many years of data. The detectability of a signal $h$ is characterized by its \emph{optimal} SNR $\rho(h;h)$, which is the SNR that would be obtained on average over noise realizations for a perfectly matching template.

The noise $S_X(f)$ includes the confusion-noise foreground $S^\mathrm{conf}_X(f)$ from unresolved sources (see Sec.\ \ref{sec:results}), as well as instrument noise $S^\mathrm{inst}_X$. For the unequal-arm Michelson observable $X$ in the (rather reasonable) limit of equal interferometer armlengths, $S^\mathrm{inst}_X$ is given by \citep{2008CQGra..25f5005V}
\begin{equation}
S^\mathrm{inst}_X = 16 \sin^2(2 \pi L f) \Bigl[2\bigl(1 + \cos^2(2 \pi L f)\bigr) S^{\Delta f/f}_\mathrm{pm}(f) + S^{\Delta f/f}_\mathrm{op}(f)\Bigr].
\end{equation}
In these units $S^{\Delta f/f}_\mathrm{pm} = S_\mathrm{pm}(f) / (2 \pi \, c \, f)$,
$S^{\Delta f/f}_\mathrm{op}(f) = S_\mathrm{op}(f) \times (2 \pi f / c)$.

To obtain the TDI response to a GW signal Eq.\ \eqref{eq:wave}, we
insert Eq.\ \eqref{eq:wave}, multiplied by the appropriate phasing
function and sky-position--dependent polarization tensors, into Eq.\
\eqref{eq:unequalmich}. The response is modulated further by the
heliocentric motion of the detector (which yields a time-dependent
Doppler shift) and its yearly rotation in a plane inclined 60$^\circ$
from the ecliptic (which generates sidebands at frequency multiples of
$1/\mathrm{yr}$).\footnote{Because ``short LISA'' orbits are very
  similar to those planned for classic LISA, this discussion is valid
  for both. Different orbits, such as halo trajectories around
  Lagrange points or Earth-centric orbits, would yield different
  signal modulations.} Both effects imprint sky-position and
polarization angles into the TDI response; this structure makes it
possible to resolve sufficiently strong GW sources that overlap in
frequency space, and to determine their approximate sky position (see,
e.g., \citealt{2004PhRvD..70b2003K}). The TDI response to a population
of  sources is given simply by the linear superposition of the
responses to individual sources.

\section{Fits of the partially subtracted confusion-noise foreground}
\label{app:subtracted}

Barack and Cutler (\citeyear{BCu04,BCu04B}) codified the lore about the \emph{total noise} of a LISA-like detector (i.e., the instrument noise plus Galactic-binary confusion noise) by providing a heuristic derivation for the well-known expression
\begin{equation}
S^\mathrm{total,BC}_h(f) = \mathrm{min} \bigl[ S_h^\mathrm{inst}(f) \times \mathrm{e}^{\kappa T^{-1} \mathrm{d}N/\mathrm{d}f}, S_h^\mathrm{inst}(f) + S_h^\mathrm{conf,BC}(f) \bigr],
\label{eq:bcone}
\end{equation}
where they estimate
\begin{equation}
S_h^\mathrm{conf,BC}(f) = 1.4 \times 10^{-44} \bigl( f / \mathrm{Hz} \bigr)^{-7/3} \, \mathrm{Hz}^{-1},
\label{eq:BCest}
\end{equation}
\begin{equation}
\frac{\mathrm{d}N}{\mathrm{d}f} = 2 \times 10^{-3} \bigl( f / \mathrm{Hz} \bigr)^{-11/3} \, \mathrm{Hz}^{-1},
\label{eq:bcthree}
\end{equation}
and $\kappa T^{-1} \simeq 1.5 / \mathrm{yr}$.
As we discuss in App.\ \ref{sec:prevwork}, under generic assumptions the \emph{unsubtracted} Galactic-binary foreground will follow a $f^{-7/3}$ spectrum;
this is confirmed in our simulations, with a coefficient $2.6 \times 10^{-44}$ for cases 1--3 and 5, and five times smaller for case 4 (see Fig.\ \ref{fig:fore}).
After the subtraction of detectable sources, the total noise found in our simulations is fit better by a smaller amplitude, a slightly different exponent, and a smoother transition function:
\begin{equation}
S^\mathrm{total,sim}_h(f) = S_h^\mathrm{inst} + S^\mathrm{conf,sim}_h \times \mathrm{tanh}^\alpha\bigl({\textstyle \frac{1}{2}} \beta \, \mathrm{d}N/\mathrm{d}f \bigr),
\label{eq:appconf}
\end{equation}
with
\begin{equation}
S^\mathrm{conf,sim}_h(f) = 1.4 \times 10^{-45} \bigl( f / \mathrm{Hz} \bigr)^{-8/3} \, \mathrm{Hz}^{-1},
\end{equation}
\begin{equation}
\frac{\mathrm{d}N}{\mathrm{d}f} = 5 \times 10^{-3} \bigl( f / \mathrm{Hz} \bigr)^{-11/3} \, \mathrm{Hz}^{-1}.
\end{equation}
For cases 1--3 and 5, and for our reference subtraction run (one year, one interferometric observable, $\rho_\mathrm{thr} = 5$), taking $\alpha = 1$, $\beta = 1/\mathrm{yr}$ yields a good fit for a 5 Mkm LISA-like detector, while $\alpha = 0.7$, $\beta = 1.2/\mathrm{yr}$ is appropriate for 1 Mkm arms. For the entire range of $\rho_\mathrm{eff}$ that we probed, good fits are obtained by setting
\begin{equation}
\begin{aligned}
& \alpha = 2 - 1.2 \, (\rho_\mathrm{eff} / 5) + 0.2 \, (\rho_\mathrm{eff} / 5)^2, \quad
& \beta & = [-0.2 + 1.5 \, (\rho_\mathrm{eff} / 5) - 0.3 \, (\rho_\mathrm{eff} / 5)^2] / \mathrm{yr} \quad && (5 \, \mathrm{Mkm}), \\ 
& \alpha = 1.6 - 1.2 \, (\rho_\mathrm{eff} / 5) + 0.3 \, (\rho_\mathrm{eff} / 5)^2, \quad
& \beta & = [0.3 + 1.3 \, (\rho_\mathrm{eff} / 5) - 0.4 \, (\rho_\mathrm{eff} / 5)^2] / \mathrm{yr} \quad && (1 \, \mathrm{Mkm}).
\end{aligned}
\label{eq:appconfend}
\end{equation}
For case 4, an acceptable fit follows from reducing $S^\mathrm{conf,sim}_h$ by a factor of five.

The equivalent confusion noise in the TDI $X$ combination (expressed as fractional-frequency fluctuations) is given by
\begin{equation}
\label{eq:applyresponse}
S^\mathrm{total,sim}_\mathrm{X}(f) = \frac{3}{20} \frac{16 \, x^2 \sin^2 x}{1 + 0.6 \, x^2} S^\mathrm{total,sim}_h(f), \;\; \mathrm{with} \;\; x = 2 \pi f L / c.
\end{equation}
Here 3/20 is the all-sky averaging $\Pfac$ discussed above; $1 + 0.6 \, x^2$ is an approximation to the frequency-dependent interferometer response function; and $16 \, x^2 \sin^2 x$ is the TDI $X$ transfer function. [In Barack and Cutler's convention (\citeyear{BCu04B}), which we follow here, the average response function is included in the definition of the equivalent strain noise $S^\mathrm{total}_h$, so that the sky-averaged SNR can be computed directly as $4 \, \mathrm{Re} \int |\tilde{h}|^2/S^\mathrm{total}_h(f)\, \mathrm{d}f$; therefore 
$S^\mathrm{conf}_h$, rather than $S^\mathrm{conf}_{h_I}$, appears in $S^\mathrm{total}_h$. By contrast, TDI observables already include the position-dependent instrument response, so it is appropriate to define the TDI noise $S^\mathrm{total}_\mathrm{X}$ without the sky averaging.]

Due to the evolving orientation of the LISA-like formation with respect to the Galaxy,
the Galactic foreground (before and after subtraction) has a strong seasonal variation in amplitude, almost constant across frequencies \citep{bibgs:giampieri+polnarev97,2004PhRvD..69l3005S,bibgs:edlund+05}. For a single TDI $X$ combination, for both our detector configurations, for all scenarios, and for our assumptions about the relative orientation of detector and Galaxy at $t = 0$ [consistent with the standard MLDC LISA orbits \citep{Arnaud:2006}], this variation is fit well by 
\begin{eqnarray}
\label{eq:timedep}
S_h^\mathrm{conf,sim}(f,t) & = & S_h^\mathrm{conf,sim}(f) \times \bigl[
1 - 0.21 \cos(2 \pi \, t/\mathrm{yr}) - 0.45 \cos(4 \pi \,
t/\mathrm{yr}) \nonumber \\ & & - 0.05 \sin(2 \pi \, t/\mathrm{yr}) + 0.01 \sin(4 \pi \, t/\mathrm{yr}) \bigr],
\end{eqnarray}
where $S_h^\mathrm{conf,sim}(f,t)$ denotes the power spectral density computed for data segments shorter than one year, but longer than the typical measurement timescale of a few hours. An equation analogous to Eq.\ \eqref{eq:timedep}, with the same coefficients, holds also for the unsubtracted background $S_h(f,t)$.

\bibliography{WDWD}

\begin{thebibliography}{100}
\expandafter\ifx\csname natexlab\endcsname\relax\def\natexlab#1{#1}\fi

\bibitem[{{Amaro-Seoane} {et~al.}(2012){Amaro-Seoane}, {Aoudia}, {Babak},
  {Bin{\'e}truy}, {Berti}, {Boh{\'e}}, {Caprini}, {Colpi}, {Cornish},
  {Danzmann}, {Dufaux}, {Gair}, {Jennrich}, {Jetzer}, {Klein}, {Lang}, {Lobo},
  {Littenberg}, {McWilliams}, {Nelemans}, {Petiteau}, {Porter}, {Schutz},
  {Sesana}, {Stebbins}, {Sumner}, {Vallisneri}, {Vitale}, {Volonteri}, \&
  {Ward}}]{esastudy}
{Amaro-Seoane}, P., {et~al.} 2012, Classical and Quantum Gravity, 29, 124016

\bibitem[{{Arnaud} {et~al.}(2006){Arnaud}, {Babak}, {Baker}, {Benacquista},
  {Cornish}, {Cutler}, {Larson}, {Sathyaprakash}, {Vallisneri}, {Vecchio}, \&
  {Vinet}}]{Arnaud:2006}
{Arnaud}, K.~A., {et~al.} 2006, in AIP Conf. Proc., Vol. 873, 625

\bibitem[{{Badenes} {et~al.}(2009){Badenes}, {Mullally}, {Thompson}, \&
  {Lupton}}]{Badenes:2009}
{Badenes}, C., {Mullally}, F., {Thompson}, S.~E., \& {Lupton}, R.~H. 2009,
  \apj, 707, 971

\bibitem[{{Barack} \& {Cutler}(2004{\natexlab{a}})}]{BCu04}
{Barack}, L., \& {Cutler}, C. 2004{\natexlab{a}}, \prd, 69, 082005

\bibitem[{{Barack} \& {Cutler}(2004{\natexlab{b}})}]{BCu04B}
---. 2004{\natexlab{b}}, \prd, 70, 122002

\bibitem[{Belczynski {et~al.}(2002)Belczynski, Kalogera, \&
  Bulik}]{bibgs:belczynski+02}
Belczynski, K., Kalogera, V., \& Bulik, T. 2002, \apj, 572, 407

\bibitem[{Belczynski {et~al.}(2008)Belczynski, Kalogera, Rasio, Taam, Zezas,
  Bulik, Maccarone, \& Ivanova}]{bibgs:belczynski+08}
Belczynski, K., Kalogera, V., Rasio, F., Taam, R., Zezas, A., Bulik, T.,
  Maccarone, T., \& Ivanova, N. 2008, \apjss, 174, 223

\bibitem[{Benacquista {et~al.}(2004)Benacquista, DeGoes, \&
  Lunder}]{bibgs:benacquista+04}
Benacquista, M., DeGoes, J., \& Lunder, D. 2004, \cqg, 21, S509

\bibitem[{Bender \& Hils(1997)}]{bibgs:bender+hils97}
Bender, P., \& Hils, D. 1997, \cqg, 14, 1439

\bibitem[{{Bildsten} {et~al.}(2007){Bildsten}, {Shen}, {Weinberg}, \&
  {Nelemans}}]{Bildsten:2007}
{Bildsten}, L., {Shen}, K.~J., {Weinberg}, N.~N., \& {Nelemans}, G. 2007,
  \apjl, 662, L95

\bibitem[{{Binney} \& {Tremaine}(1987)}]{BinneyTremaine:1987}
{Binney}, J., \& {Tremaine}, S. 1987, {Galactic dynamics}

\bibitem[{{Boissier} \& {Prantzos}(1999)}]{Boissier:1999}
{Boissier}, S., \& {Prantzos}, N. 1999, \mnras, 307, 857

\bibitem[{{Bovy} {et~al.}(2012){Bovy}, {Rix}, \& {Hogg}}]{Bovy:2011}
{Bovy}, J., {Rix}, H.-W., \& {Hogg}, D.~W. 2012, \apj, 751, 131

\bibitem[{{Brown} {et~al.}(2010){Brown}, {Kilic}, {Allende Prieto}, \&
  {Kenyon}}]{Brown:2010}
{Brown}, W.~R., {Kilic}, M., {Allende Prieto}, C., \& {Kenyon}, S.~J. 2010,
  \apj, 723, 1072

\bibitem[{{Brown} {et~al.}(2011){Brown}, {Kilic}, {Allende Prieto}, \&
  {Kenyon}}]{Brown:2011}
---. 2011, \mnras, 411, L31

\bibitem[{{Brown} {et~al.}(2012){Brown}, {Kilic}, {Allende Prieto}, \&
  {Kenyon}}]{Brown:2012}
---. 2012, \apj, 744, 142

\bibitem[{{Cornish} \& {Littenberg}(2007)}]{Cornish:2007}
{Cornish}, N.~J., \& {Littenberg}, T.~B. 2007, \prd, 76, 083006

\bibitem[{{Crowder} \& {Cornish}(2004)}]{Crowder:2004}
{Crowder}, J., \& {Cornish}, N.~J. 2004, \prd, 70, 082004

\bibitem[{{Cutler} \& {Flanagan}(1994)}]{1994PhRvD..49.2658C}
{Cutler}, C., \& {Flanagan}, {\'E}.~E. 1994, \prd, 49, 2658

\bibitem[{{Dan} {et~al.}(2011){Dan}, {Rosswog}, {Guillochon}, \&
  {Ramirez-Ruiz}}]{Dan:2011}
{Dan}, M., {Rosswog}, S., {Guillochon}, J., \& {Ramirez-Ruiz}, E. 2011, \apj,
  737, 89

\bibitem[{{Dan} {et~al.}(2012){Dan}, {Rosswog}, {Guillochon}, \&
  {Ramirez-Ruiz}}]{Dan:2012}
---. 2012, \mnras, 422, 2417

\bibitem[{Edlund {et~al.}(2005)Edlund, Tinto, Kr{\'o}lak, \&
  Nelemans}]{bibgs:edlund+05}
Edlund, J., Tinto, M., Kr{\'o}lak, A., \& Nelemans, G. 2005, \prd, 71, 122003

\bibitem[{{Ergma} \& {Fedorova}(1990)}]{Ergma:1990}
{Ergma}, E.~V., \& {Fedorova}, A.~V. 1990, \apss, 163, 143

\bibitem[{{Estabrook} \& {Wahlquist}(1975)}]{1975GReGr...6..439E}
{Estabrook}, F.~B., \& {Wahlquist}, H.~D. 1975, Gen. Rel. Grav., 6, 439

\bibitem[{{Evans} {et~al.}(1987){Evans}, {Iben}, \& {Smarr}}]{EIS87}
{Evans}, C.~R., {Iben}, Jr., I., \& {Smarr}, L. 1987, \apj, 323, 129

\bibitem[{Farmer \& Phinney(2003)}]{bibgs:farmer+phinney03}
Farmer, A., \& Phinney, E. 2003, \mnras, 346, 1197

\bibitem[{{Farmer} \& {Roelofs}(2010)}]{Farmer:2010}
{Farmer}, A., \& {Roelofs}, G. 2010, arXiv:1006.4112

\bibitem[{Fontaine {et~al.}(2011)Fontaine, Brassard, Green, Charpinet, Dufour,
  Hubeny, Steeghs, Aerts, Randall, Bergeron, Guvenen, O'Malley, Grootel,
  ÿstensen, Bloemen, Silvotti, Howell, Baran, Kepler, Marsh, Montgomery,
  Oreiro, Provencal, Telting, Winget, Zima, Christensen-Dalsgaard, \&
  Kjeldsen}]{Fontaine:2011}
Fontaine, G., {et~al.} 2011, \apj, 726, 92

\bibitem[{Giampieri \& Polnarev(1997)}]{bibgs:giampieri+polnarev97}
Giampieri, G., \& Polnarev, A. 1997, \mnras, 291, 149

\bibitem[{{Gokhale} {et~al.}(2007){Gokhale}, {Peng}, \& {Frank}}]{Gokhale:2007}
{Gokhale}, V., {Peng}, X.~M., \& {Frank}, J. 2007, \apj, 655, 1010

\bibitem[{Han(1998)}]{bibgs:han98}
Han, Z. 1998, \mnras, 296, 1019

\bibitem[{Hils(1998)}]{bibgs:hils98}
Hils, D. 1998, in AIP Conf. Proc., Vol. 456, 68

\bibitem[{Hils {et~al.}(1990)Hils, Bender, \& Webbink}]{bibgs:hils+90}
Hils, D., Bender, P., \& Webbink, R. 1990, \apj, 360, 75

\bibitem[{{Hils} \& {Bender}(2000)}]{HB00}
{Hils}, D., \& {Bender}, P.~L. 2000, \apj, 537, 334

\bibitem[{{Holberg} {et~al.}(2008){Holberg}, {Sion}, {Oswalt}, {McCook},
  {Foran}, \& {Subasavage}}]{Holberg:2008}
{Holberg}, J.~B., {Sion}, E.~M., {Oswalt}, T., {McCook}, G.~P., {Foran}, S., \&
  {Subasavage}, J.~P. 2008, \aj, 135, 1225

\bibitem[{Hughes(2002)}]{bibgs:hughes02}
Hughes, S. 2002, \mnras, 331, 805

\bibitem[{Hurley {et~al.}(2000)Hurley, Pols, \& Tout}]{bibgs:hurley+00}
Hurley, J., Pols, O., \& Tout, C. 2000, \mnras, 315, 543

\bibitem[{Hurley {et~al.}(2002)Hurley, Tout, \& Pols}]{bibgs:hurley+02}
Hurley, J., Tout, C., \& Pols, O. 2002, \mnras, 329, 897

\bibitem[{{Iben} \& {Tutukov}(1984)}]{Iben:1984}
{Iben}, Jr., I., \& {Tutukov}, A.~V. 1984, \apjs, 54, 335

\bibitem[{Jennrich(2009)}]{Jennrich:2009p1398}
Jennrich, O. 2009, \cqg, 26, 153001

\bibitem[{{Juri{\'c}} {et~al.}(2008){Juri{\'c}}, {Ivezi{\'c}}, {Brooks},
  {Lupton}, {Schlegel}, {Finkbeiner}, {Padmanabhan}, {Bond}, {Sesar},
  {Rockosi}, {Knapp}, {Gunn}, {Sumi}, {Schneider}, {Barentine}, {Brewington},
  {Brinkmann}, {Fukugita}, {Harvanek}, {Kleinman}, {Krzesinski}, {Long},
  {Neilsen}, {Nitta}, {Snedden}, \& {York}}]{Juric:2008}
{Juri{\'c}}, M., {et~al.} 2008, \apj, 673, 864

\bibitem[{{Kilic} {et~al.}(2011{\natexlab{a}}){Kilic}, {Brown}, {Allende
  Prieto}, {Ag{\"u}eros}, {Heinke}, \& {Kenyon}}]{Kilic:2011}
{Kilic}, M., {Brown}, W.~R., {Allende Prieto}, C., {Ag{\"u}eros}, M.~A.,
  {Heinke}, C., \& {Kenyon}, S.~J. 2011{\natexlab{a}}, \apj, 727, 3

\bibitem[{{Kilic} {et~al.}(2011{\natexlab{b}}){Kilic}, {Brown}, {Hermes},
  {Allende Prieto}, {Kenyon}, {Winget}, \& {Winget}}]{Kilic:2011b}
{Kilic}, M., {Brown}, W.~R., {Hermes}, J.~J., {Allende Prieto}, C., {Kenyon},
  S.~J., {Winget}, D.~E., \& {Winget}, K.~I. 2011{\natexlab{b}}, \mnras, 418,
  L157

\bibitem[{{Kr{\'o}lak} {et~al.}(2004){Kr{\'o}lak}, {Tinto}, \&
  {Vallisneri}}]{2004PhRvD..70b2003K}
{Kr{\'o}lak}, A., {Tinto}, M., \& {Vallisneri}, M. 2004, \prd, 70, 022003

\bibitem[{{Kroupa} {et~al.}(1993){Kroupa}, {Tout}, \& {Gilmore}}]{Kroupa:1993}
{Kroupa}, P., {Tout}, C.~A., \& {Gilmore}, G. 1993, \mnras, 262, 545

\bibitem[{{Kulkarni} \& {van Kerkwijk}(2010)}]{Kulkarni:2010}
{Kulkarni}, S.~R., \& {van Kerkwijk}, M.~H. 2010, \apj, 719, 1123

\bibitem[{{Levitan} {et~al.}(2011){Levitan}, {Fulton}, {Groot}, {Kulkarni},
  {Ofek}, {Prince}, {Shporer}, {Bloom}, {Cenko}, {Kasliwal}, {Law}, {Nugent},
  {Poznanski}, {Quimby}, {Horesh}, {Sesar}, \& {Sternberg}}]{Levitan:2011}
{Levitan}, D., {et~al.} 2011, \apj, 739, 68

\bibitem[{Lipunov {et~al.}(1995)Lipunov, Nazin, Panchenko, Postnov, \&
  Prokhorov}]{bibgs:lipunov+95}
Lipunov, V., Nazin, S., Panchenko, I., Postnov, K., \& Prokhorov, M. 1995, \aa,
  298, 677

\bibitem[{Lipunov \& Postnov(1987)}]{bibgs:lipunov+87a}
Lipunov, V., \& Postnov, K. 1987, Astronomicheskii Zhurnal, 64, 438

\bibitem[{Lipunov {et~al.}(1987)Lipunov, Postnov, \&
  Prokhorov}]{bibgs:lipunov+87b}
Lipunov, V., Postnov, K., \& Prokhorov, M. 1987, \aa, 176, L1

\bibitem[{Lipunov {et~al.}(1996)Lipunov, Postnov, \&
  Prokhorov}]{bibgs:lipunov+96}
---. 1996, \aa, 310, 489

\bibitem[{Liu(2009)}]{bibgs:liu09}
Liu, J. 2009, \mnras, 400, 1850

\bibitem[{Liu {et~al.}(2010)Liu, Zhang, Han, \& Zhang}]{bibgs:liu+10a}
Liu, J., Zhang, Y., Han, Z., \& Zhang, F. 2010, Astrophysics and Space Science,
  329, 297

\bibitem[{Maggiore(2007)}]{Maggiore:1900zz}
Maggiore, M. 2007, {Gravitational Waves. Vol. 1: Theory and Experiments}
  (Oxford, UK: Oxford University Press)

\bibitem[{{Marsh}(2011)}]{Marsh:2011}
{Marsh}, T.~R. 2011, \cqg, 28, 094019

\bibitem[{{Marsh} {et~al.}(2011){Marsh}, {G{\"a}nsicke}, {Steeghs},
  {Southworth}, {Koester}, {Harris}, \& {Merry}}]{Marsh:2010}
{Marsh}, T.~R., {G{\"a}nsicke}, B.~T., {Steeghs}, D., {Southworth}, J.,
  {Koester}, D., {Harris}, V., \& {Merry}, L. 2011, \apj, 736, 95

\bibitem[{{Marsh} {et~al.}(2004){Marsh}, {Nelemans}, \& {Steeghs}}]{Marsh:2004}
{Marsh}, T.~R., {Nelemans}, G., \& {Steeghs}, D. 2004, \mnras, 350, 113

\bibitem[{{Marsh} \& {Steeghs}(2002)}]{MarshSteeghs:2002}
{Marsh}, T.~R., \& {Steeghs}, D. 2002, \mnras, 331, L7

\bibitem[{{Maxted} \& {Marsh}(1999)}]{Maxted:1999}
{Maxted}, P.~F.~L., \& {Marsh}, T.~R. 1999, \mnras, 307, 122

\bibitem[{{Mullally} {et~al.}(2009){Mullally}, {Badenes}, {Thompson}, \&
  {Lupton}}]{Mullally:2009}
{Mullally}, F., {Badenes}, C., {Thompson}, S.~E., \& {Lupton}, R. 2009, \apjl,
  707, L51

\bibitem[{{Napiwotzki}(2009)}]{Napiwotzki:2009}
{Napiwotzki}, R. 2009, J. Phys. Conf. Series, 172, 012004

\bibitem[{{Nelemans}(2009)}]{Nelemans:2009}
{Nelemans}, G. 2009, Classical and Quantum Gravity, 26, 094030

\bibitem[{Nelemans(2012)}]{Nelemans}
Nelemans, G. 2012, LISA wiki (verification binaries):
  \url{astro.ru.nl/~nelemans/dokuwiki/doku.php?id=verification_binaries:intro}

\bibitem[{{Nelemans} {et~al.}(2001{\natexlab{a}}){Nelemans}, {Portegies Zwart},
  {Verbunt}, \& {Yungelson}}]{NPVY01}
{Nelemans}, G., {Portegies Zwart}, S.~F., {Verbunt}, F., \& {Yungelson}, L.~R.
  2001{\natexlab{a}}, \aap, 368, 939

\bibitem[{{Nelemans} {et~al.}(2001{\natexlab{b}}){Nelemans}, {Yungelson}, \&
  {Portegies Zwart}}]{NYP01}
{Nelemans}, G., {Yungelson}, L.~R., \& {Portegies Zwart}, S.~F.
  2001{\natexlab{b}}, \aap, 375, 890

\bibitem[{{Nelemans} {et~al.}(2004){Nelemans}, {Yungelson}, \& {Portegies
  Zwart}}]{Nelemans:2004}
---. 2004, \mnras, 349, 181

\bibitem[{{Nelemans} {et~al.}(2001{\natexlab{c}}){Nelemans}, {Yungelson},
  {Portegies Zwart}, \& {Verbunt}}]{Nelemans2001ddwd}
{Nelemans}, G., {Yungelson}, L.~R., {Portegies Zwart}, S.~F., \& {Verbunt}, F.
  2001{\natexlab{c}}, \aap, 365, 491

\bibitem[{Peters \& Mathews(1963)}]{Peters:1963ux}
Peters, P., \& Mathews, J. 1963, Phys. Rev., 131, 435

\bibitem[{{Piro}(2011)}]{Piro:2011}
{Piro}, A.~L. 2011, \apjl, 740, L53

\bibitem[{{Podsiadlowski} {et~al.}(2001){Podsiadlowski}, {Han}, \&
  {Rappaport}}]{Podsiadlowski:2001}
{Podsiadlowski}, P., {Han}, Z., \& {Rappaport}, S. 2001, astro-ph/0109171

\bibitem[{Portegies~Zwart \& Yungelson(1998)}]{bibgs:zwart+98}
Portegies~Zwart, S., \& Yungelson, L. 1998, \aa, 332, 173

\bibitem[{{Portegies Zwart} \& {Verbunt}(1996)}]{PortegiesZwart:1996}
{Portegies Zwart}, S.~F., \& {Verbunt}, F. 1996, \aap, 309, 179

\bibitem[{Postnov \& Prokhorov(1998)}]{bibgs:postnov+prokhorov98}
Postnov, K., \& Prokhorov, M. 1998, \apj, 494, 674

\bibitem[{Prince {et~al.}(2009)}]{lisasciencecase}
Prince, T.~A., {et~al.} 2009, {LISA: Probing the Universe with Gravitational
  Waves}, \url{list.caltech.edu/mission_documents}

\bibitem[{{Ramsay} {et~al.}(2002){Ramsay}, {Hakala}, \&
  {Cropper}}]{Ramsay:2002}
{Ramsay}, G., {Hakala}, P., \& {Cropper}, M. 2002, \mnras, 332, L7

\bibitem[{{Rau} {et~al.}(2010){Rau}, {Roelofs}, {Groot}, {Marsh}, {Nelemans},
  {Steeghs}, {Salvato}, \& {Kasliwal}}]{Rau:2010}
{Rau}, A., {Roelofs}, G.~H.~A., {Groot}, P.~J., {Marsh}, T.~R., {Nelemans}, G.,
  {Steeghs}, D., {Salvato}, M., \& {Kasliwal}, M.~M. 2010, \apj, 708, 456

\bibitem[{{Roelofs} {et~al.}(2007{\natexlab{a}}){Roelofs}, {Groot}, {Benedict},
  {McArthur}, {Steeghs}, {Morales-Rueda}, {Marsh}, \&
  {Nelemans}}]{Roelofs:2007b}
{Roelofs}, G.~H.~A., {Groot}, P.~J., {Benedict}, G.~F., {McArthur}, B.~E.,
  {Steeghs}, D., {Morales-Rueda}, L., {Marsh}, T.~R., \& {Nelemans}, G.
  2007{\natexlab{a}}, \apj, 666, 1174

\bibitem[{{Roelofs} {et~al.}(2007{\natexlab{b}}){Roelofs}, {Nelemans}, \&
  {Groot}}]{Roelofs:2007}
{Roelofs}, G.~H.~A., {Nelemans}, G., \& {Groot}, P.~J. 2007{\natexlab{b}},
  \mnras, 382, 685

\bibitem[{{Ruiter} {et~al.}(2009){Ruiter}, {Belczynski}, {Benacquista}, \&
  {Holley-Bockelmann}}]{Ruiter:2009}
{Ruiter}, A.~J., {Belczynski}, K., {Benacquista}, M., \& {Holley-Bockelmann},
  K. 2009, \apj, 693, 383

\bibitem[{{Ruiter} {et~al.}(2010){Ruiter}, {Belczynski}, {Benacquista},
  {Larson}, \& {Williams}}]{Ruiter:2010}
{Ruiter}, A.~J., {Belczynski}, K., {Benacquista}, M., {Larson}, S.~L., \&
  {Williams}, G. 2010, \apj, 717, 1006

\bibitem[{{Scannapieco} {et~al.}(2011){Scannapieco}, {White}, {Springel}, \&
  {Tissera}}]{Scannapieco:2011}
{Scannapieco}, C., {White}, S.~D.~M., {Springel}, V., \& {Tissera}, P.~B. 2011,
  \mnras, 417, 154

\bibitem[{{Sch{\"o}nrich} \& {Binney}(2009{\natexlab{a}})}]{Schonrich:2009a}
{Sch{\"o}nrich}, R., \& {Binney}, J. 2009{\natexlab{a}}, \mnras, 396, 203

\bibitem[{{Sch{\"o}nrich} \& {Binney}(2009{\natexlab{b}})}]{Schonrich:2009b}
---. 2009{\natexlab{b}}, \mnras, 399, 1145

\bibitem[{Seto(2002)}]{bibgs:seto02}
Seto, N. 2002, \mnras, 333, 469

\bibitem[{{Seto}(2004)}]{2004PhRvD..69l3005S}
{Seto}, N. 2004, \prd, 69, 123005

\bibitem[{{Shen} {et~al.}(2010){Shen}, {Kasen}, {Weinberg}, {Bildsten}, \&
  {Scannapieco}}]{Shen:2010}
{Shen}, K.~J., {Kasen}, D., {Weinberg}, N.~N., {Bildsten}, L., \&
  {Scannapieco}, E. 2010, \apj, 715, 767

\bibitem[{{Siebert} {et~al.}(2011){Siebert}, {Williams}, {Siviero}, {Reid},
  {Boeche}, {Steinmetz}, {Fulbright}, {Munari}, {Zwitter}, {Watson}, {Wyse},
  {de Jong}, {Enke}, {Anguiano}, {Burton}, {Cass}, {Fiegert}, {Hartley},
  {Ritter}, {Russel}, {Stupar}, {Bienaym{\'e}}, {Freeman}, {Gilmore}, {Grebel},
  {Helmi}, {Navarro}, {Binney}, {Bland-Hawthorn}, {Campbell}, {Famaey},
  {Gerhard}, {Gibson}, {Matijevi{\v c}}, {Parker}, {Seabroke}, {Sharma},
  {Smith}, \& {Wylie-de Boer}}]{Siebert:2011}
{Siebert}, A., {et~al.} 2011, \aj, 141, 187

\bibitem[{{Sofue} {et~al.}(2009){Sofue}, {Honma}, \& {Omodaka}}]{Sofue:2009}
{Sofue}, Y., {Honma}, M., \& {Omodaka}, T. 2009, \pasj, 61, 227

\bibitem[{{Solheim}(2010)}]{Solheim:2010}
{Solheim}, J.-E. 2010, \pasp, 122, 1133

\bibitem[{{Timpano} {et~al.}(2006){Timpano}, {Rubbo}, \&
  {Cornish}}]{Timpano:2006}
{Timpano}, S.~E., {Rubbo}, L.~J., \& {Cornish}, N.~J. 2006, \prd, 73, 122001

\bibitem[{{Toonen} {et~al.}(2011){Toonen}, {Nelemans}, \& {Portegies
  Zwart}}]{Toonen:2011}
{Toonen}, S., {Nelemans}, G., \& {Portegies Zwart}, S. 2011, arXiv:1101.2787

\bibitem[{Tutukov \& Yungelson(1996)}]{bibgs:tutukov+yungelson96}
Tutukov, A., \& Yungelson, L. 1996, \mnras, 280, 1035

\bibitem[{{Tutukov} \& {Fedorova}(1989)}]{Tutukov:1989}
{Tutukov}, A.~V., \& {Fedorova}, A.~V. 1989, \sovast, 33, 606

\bibitem[{{Vallisneri}(2005)}]{2005PhRvD..72d2003V}
{Vallisneri}, M. 2005, \prd, 72, 042003

\bibitem[{Vallisneri(2011)}]{Vallisneri:2011}
Vallisneri, M. 2011, package fastbinary, lisasolve.googlecode.com, rev458

\bibitem[{{Vallisneri} {et~al.}(2008){Vallisneri}, {Crowder}, \&
  {Tinto}}]{2008CQGra..25f5005V}
{Vallisneri}, M., {Crowder}, J., \& {Tinto}, M. 2008, \cqg, 25, 065005

\bibitem[{{Verbunt} \& {Rappaport}(1988)}]{Verbunt:1988}
{Verbunt}, F., \& {Rappaport}, S. 1988, \apj, 332, 193

\bibitem[{Webbink \& Han(1998)}]{bibgs:webbink+han98}
Webbink, R., \& Han, Z. 1998, in AIP Conf. Proc., Vol. 456, 61

\bibitem[{{Webbink}(1984)}]{Webbink:1984}
{Webbink}, R.~F. 1984, \apj, 277, 355

\bibitem[{{Yu} \& {Jeffery}(2010)}]{Yu:2010}
{Yu}, S., \& {Jeffery}, C.~S. 2010, \aap, 521, A85

\end{thebibliography}

\end{document}